\documentclass[prl,reprint,twocolumn,superscriptaddress,noshowpacs,notitlepage,longbibliography]{revtex4-1}%
\usepackage{graphicx,bm,times}
\usepackage{amsmath}
\usepackage{amsfonts}
\usepackage{amssymb}
\usepackage{color}
\usepackage{hyperref}
\hypersetup{
	colorlinks = true,
	allcolors = {blue}
}

\begin{document}
	
	\title{The three-dimensional electronic structure of the nematic and antiferromagnetic phases of NaFeAs from detwinned ARPES measurements}
	
	\author{Matthew D. Watson}
	\email[corresponding author:]{mdw5@st-andrews.ac.uk}
	\affiliation{Diamond Light Source, Harwell Campus, Didcot, OX11 0DE, United Kingdom}
	\affiliation{School of Physics and Astronomy, University of St. Andrews, St. Andrews KY16 9SS, United Kingdom}

	\author{Saicharan Aswartham}
	\affiliation{Leibniz Institute for Solid State and Materials Research, 01171 Dresden, Germany}
			
	\author{Luke C. Rhodes}
	\affiliation{Diamond Light Source, Harwell Campus, Didcot, OX11 0DE, United Kingdom}
	\affiliation{Department of Physics, Royal Holloway, University of London, Egham, Surrey, TW20 0EX, United Kingdom}
	
	\author{Benjamin Parrett}
	\affiliation{Diamond Light Source, Harwell Campus, Didcot, OX11 0DE, United Kingdom}
	\affiliation{London Centre for Nanotechnology, Gordon Street, London, WC1H 0AH, United Kingdom}
	
	\author{Hideaki Iwasawa}
	\affiliation{Diamond Light Source, Harwell Campus, Didcot, OX11 0DE, United Kingdom}

	\author{Moritz Hoesch}
	\affiliation{Diamond Light Source, Harwell Campus, Didcot, OX11 0DE, United Kingdom}
	
	\author{Igor Morozov}
	\affiliation{Leibniz Institute for Solid State and Materials Research, 01171 Dresden, Germany}
	\affiliation{Lomonosov Moscow State University, 119991 Moscow, Russia.}
		
	\author{Bernd B\"{u}chner}
	\affiliation{Leibniz Institute for Solid State and Materials Research, 01171 Dresden, Germany}
	
	\author{Timur K. Kim}
	\email[corresponding author:]{timur.kim@diamond.ac.uk}
	\affiliation{Diamond Light Source, Harwell Campus, Didcot, OX11 0DE, United Kingdom}

	\begin{abstract}
		
		We report a comprehensive ARPES study of NaFeAs, a prototypical parent compound of the Fe-based superconductors. By mechanically detwinning the samples, we show that in the nematic phase (below the structural transition at $T_s$ = 54 K but above the antiferromagnetic transition at $T_N$ = 43 K) spectral weight is detected on only the elliptical electron pocket along the longer $a_{orth}$ axis. This dramatic anisotropy is likely to arise as a result of coupling to a fluctuating antiferromagnetic order in the nematic phase. In the long-range ordered antiferromagnetic state below $T_N$, this single electron pocket is backfolded and hybridises with the hole bands, leading to the reconstructed Fermi surface. By careful analysis of the $k_z$ variation, we show that the backfolding of spectral weight in the magnetic phase has a wavector of ($\pi$,0,$\pi$), with the $c$-axis component being in agreement with the magnetic ordering in NaFeAs observed by neutron scattering. Our results clarify the origin of the tiny Fermi surfaces of NaFeAs at low temperatures and highlight the importance of the three-dimensional aspects of the electronic and magnetic properties of Fe-based superconductors.   
		
	\end{abstract}
	%\date{\today}
	\maketitle
	
	%%%%%%%%%%%%%%%%%%%%%%%%%%%%%%%%%%%%%%%%%%%%%%%%

The dome of unconventional superconductivity in the phase diagrams of Fe-based superconductors usually appears once the stripe antiferromagnetic order found in the parent compounds is suppressed. This proximity has inspired many works which use a spin-fluctuation pairing mechanism to explain the relatively high $T_c$ superconductivity in these materials \cite{Mazin2008,Hirschfeld2011}. However this picture is challenged by the existence of a distinct ``nematic" phase in compounds such as NaFeAs, characterised by a tetragonal-orthorhombic structural transition of the lattice and twofold symmetry of the electronic structure, but without static magnetic order \cite{Li2009,Presniakov2013}. This led to some suggestions that there is a separate symmetry-breaking instability of orbital order \cite{Kontani2010}, which was supported by an apparently large energy scale for $d_{xz}-d_{yz}$ orbital splitting in detwinned ARPES measurements of BaFe$_2$As$_2$ \cite{Yi2011} and NaFeAs \cite{Yi2012,Zhang2012b}. However the origin of the nematic phase is still not settled, with other groups proposing a magnetically-driven `spin-nematic' state characterised by strong antiferromagnetic fluctuations with a chosen direction but finite magnetic correlation length \cite{Wright2012,Fernandes2014,Rosenthal2014}. Thus it is important to continue to look for new insights in the parent compounds of the Fe-based superconductors, especially in systems where the nematic and magnetic phases can be distinguished such as NaFeAs.

NaFeAs is a prototypical parent compound of the Fe-based superconductors, with a tetragonal-orthorhombic structural transition at $T_s$ = 54 K, and a magnetic transition at $T_N$ = 43 K. A superconducting dome reaching a maximum $T_c \approx $ 20 K is found upon doping with Co \cite{Wright2012,Deng2015}, Ni \cite{Parker2010} or Rh \cite{Steckel2015} on the Fe site. While not exhibiting ordering temperatures as high as BaFe$_2$As$_2$ ($T_s \approx T_N \approx$ 134 K \cite{Kim2011}) or LaFeAsO ($T_s \approx$ 165~K, $T_N \approx$ 145~K \cite{Zhang2015PRL}), it offers a convenient test bed for ideas about the nematic phase due to the relatively large temperature separation of $T_s$ and $T_N$. Moreover the system is well-suited for ARPES measurements as large and high quality single crystals are available that may be reliably cleaved to yield flat non-polar surfaces, similar to its well-studied sister compound LiFeAs \cite{Borisenko2010,Borisenko2015}. Previous ARPES studies of detwinned NaFeAs samples focused on the different dispersions observed close to the $\rm \bar{M}_X$ and $\rm \bar{M}_Y$ points below $T_s$ \cite{Yi2012,Zhang2012b}. This was interpreted as evidence for a $d_{xz}-d_{yz}$ orbital splitting on a large $\sim$30-40 meV energy scale, similar to pioneering measurements of detwinned BaFe$_2$As$_2$ \cite{Yi2011}. For some time, this apparent orbital splitting was considered to be a hallmark of the nematic phase \cite{Yi2017_arxiv}. However recent high-resolution detwinned measurements of FeSe were interpreted in quite a different scheme \cite{Watson2017d_arxiv}, making it important to re-examine whether the famous $d_{xz}-d_{yz}$ orbital splitting around the $\rm \bar{M}$ points was really a robust conclusion in NaFeAs.    

In this paper we report extensive ARPES measurements of twinned and detwinned NaFeAs. By detwinning the samples we show that the most distinctive feature of the nematic phase is that despite the expectation of observing two crossed elliptical electron pockets expected at the M point, ARPES measurements only find spectral weight on the elliptical electron pocket which is directed along the longer $a_{orth}$ axis of the orthorhombic structure. Since this dramatic anisotropy of the spectral weight cannot be explained by any orbital order parameter, we attribute it to a coupling with strong but not static antiferromagnetic fluctuations in the nematic phase. Correspondingly, the different dispersions observed along $\Gamma$-$\rm M_X$ and $\Gamma$-$\rm M_Y$ are interpreted as a consequence of this ``one-ellipse" structure, and are not a signature of $d_{xz}-d_{yz}$ orbital polarisation as was previously thought. In the magnetic phase, we show that this single electron-like ellipse is backfolded and hybridises with the hole bands, leading to the reconstructed Fermi surface, which includes tiny pockets with Dirac-like dispersions. By careful analysis of the $k_z$ variation of the pockets, we show that the magnetic backfolding has a wavector of ($\pi$,0,$\pi$), with the $c$-axis component being in agreement with the magnetic ordering in NaFeAs inferred from neutron scattering measurements \cite{Li2009}. Our results hint indirectly at a strong role for antiferromagnetic correlations in the nematic phase, show the origin of the tiny Fermi surfaces of NaFeAs at low temperatures, and emphasize the importance of considering the $k_z$-dependence of the electronic and magnetic structures of Fe-based superconductors.

\begin{figure*}[t!]
	\centering
	\includegraphics[width=\linewidth]{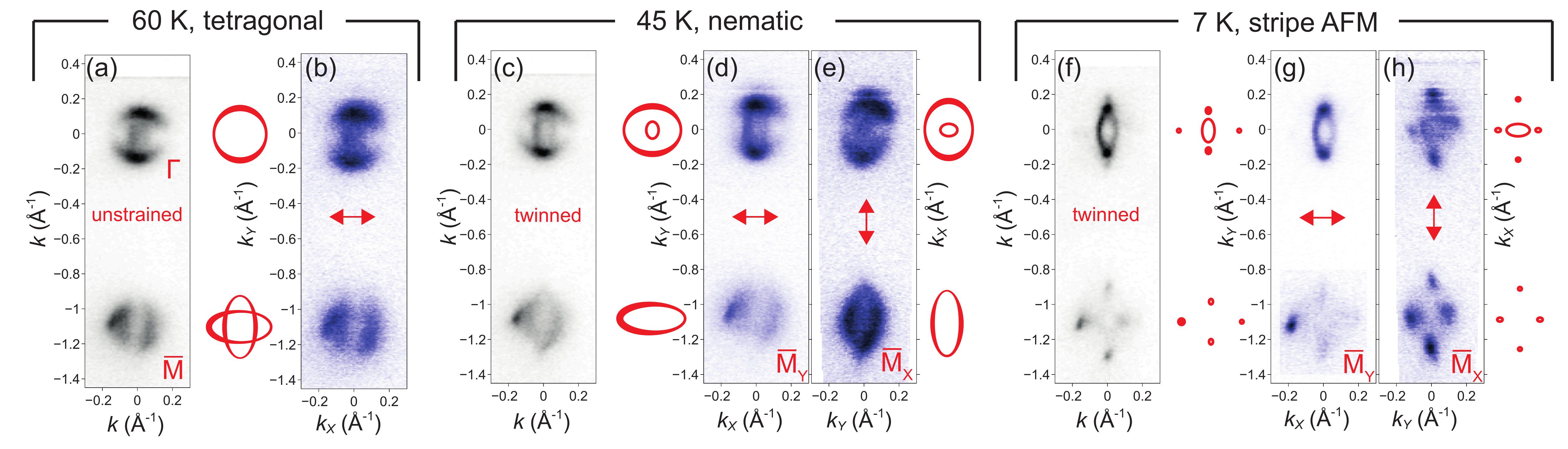}
	\caption{Fermi surface maps obtained at 42 eV in LH polarisation of (a) unstrained and (b) strained NaFeAs samples at 60~K in the tetragonal phase. Red double-headed arrows indicate the direction of the uniaxial tensile strain on the sample, which will correspond to the longer $a_{orth}$ direction in the nematic and magnetic phases. Schematic Fermi surfaces are drawn in red, with the line thickness representing the photoemission intensity in this particular measurement geometry (see also SM). The schematic Fermi surfaces also take into account data obtained in LV polarisation (not shown here, but see Fig.~\ref{fig:fig-backfolding} and SM), where some of the bands are better observed due to selection rules.  (c-e) Equivalent measurements at 45 K in the nematic phase and (f-h) at 7 K in the magnetic phase.}
	\label{fig:fig-detwinned-1}
\end{figure*}

\section{Methods}

Samples were grown by the method reported in Ref.~\cite{Steckel2015}. ARPES measurements were performed at the I05 beamline at the Diamond Light Source \cite{I05beamlinepaper} using linear horizontal (LH) and vertical (LV) polarisations. The photoelectron energy and angular distributions were analyzed with a SCIENTA R4000 hemispherical analyzer. The sample remained in a vacuum of $<2\times10^{-10}$ mbar throughout the measurements. The angular resolution was 0.2$^\circ$, and the overall energy resolution was better than 10 meV in all measurements. Since NaFeAs is a very air-sensitive material, samples were prepared in a dry argon glove box and transferred into the measurement chamber without any exposure to air.  
In this paper we label high-symmetry points in the Brillouin zone by their formal labels in the crystallographic 2-Fe unit cell in the tetragonal phase. The direction of the tensile strain determines the orientation of the longer $a_{orth}$ axis. In the detwinned measurements in the nematic or magnetic phases we distinguish the M points along the (antiferromagnetic) $a_{orth}$ and (ferromagnetic) $b_{orth}$ axes as $\rm M_X$ and $\rm M_Y$ respectively, while we use only $\rm M$ for twinned measurements. Similarly, the axis labels $k_x$ and $k_y$ are defined with respect to the $a_{orth}$,$b_{orth}$ axes, while for twinned data we use $k$. In some data sets where the photon energy does not correspond to a high-symmetry point in $k_z$ we use the notation $\rm \bar{M}_{X,Y}$. The 1-Fe unit cell notation is used for wavevectors. 

The parent phases of Fe-based superconductors naturally form orthorhombic twin domains when cooled below $T_s$. Since the size of the domains is usually much smaller than the beam spot, a normal ARPES measurement would observe a superposition of spectra from the two domains \footnote{Formally there are four possible domains in the magnetic phase but for the purposes of ARPES measurements we can simply consider only two possible domain orientations.}. However it is known that the application of a modest strain along the Fe-Fe direction can promote the volume fractions of one of the domains \cite{Fisher2011a}, and thus it is possible to effectively probe the electronic structure of a single domain by measuring samples under strain \footnote{The detwinned data have slightly compromised resolution compared with the unstrained sample; this is partially due to necessarily shorter measurement times on the detwinned samples and possibly extrinsic issues due to the more complex sample mounting environment}. In our experiment, samples are put under tensile strain on a horseshoe-shaped device described in Ref.~\cite{Watson2017d_arxiv}.

\section{Results}

In Fig.~\ref{fig:fig-detwinned-1}(a,b) we present ARPES measurements of NaFeAs with and without a uniaxial strain applied to the sample. For the measurements at 60 K at the $\Gamma$ point, the Fermi surface appears to consist of a single hole pocket, with a second ``incipient" \cite{Chen2015} band with a maximum just below the Fermi level. The observed Fermi surfaces are substantially smaller than the predictions of DFT, which is commonly observed in Fe-based superconductors \cite{Borisenko2015} and may be related to significant non-local interactions \cite{Jiang2016} or inter-band scattering \cite{Ortenzi2009}. In addition, the dispersions also have significant bandwidth renormalisations compared with DFT predictions (e.g. see Fig.~\ref{fig:fig-kz}) and a crossover to incoherent excitations at higher binding energies has been previously reported \cite{Charnukha2016,Evtushinsky2017NaFeAs,Nekrasov2015},  similar to results in FeSe \cite{Evtushinsky2017_arxiv,Watson2017c}. All of this points to the influence of strong electronic correlations in shaping the overall spectral function of NaFeAs, though here we focus on the evolution of the quasiparticle dispersions.  

The measurements of the electron pockets are strongly modulated by non-trivial matrix element effects \cite{Brouet2012}. Although there has been some controversy, the highest-resolution recent results in LiFeAs \cite{Borisenko2015}, FeSe in the tetragonal phase \cite{Watson2016}, and slightly overdoped tetragonal NaFe$_{0.978}$Co$_{0.022}$As \cite{Charnukha2016} have shown that, at least at some optimal photon energies and geometries, ARPES measurements in the tetragonal phase can observe a Fermi surface consisting of two crossed ellipses, corresponding to the expected band structure of the 2-Fe unit cell. However in many geometries and incident beam conditions only one of the ellipses dominates the photoemission intensity, as is the case in the 60~K Fermi surface maps presented in Fig.~\ref{fig:fig-detwinned-1}(a,b), where the sides of the vertical ellipse are observed most clearly. Nevertheless there is an additional arc-shaped feature on the left hand side of the Fermi surface map which is a contribution from the horizontal ellipse, as indicated schematically.  

At 60~K the applied strain on the sample in Fig.~\ref{fig:fig-detwinned-1}(b) appears to have little effect, compared with an unstrained sample in Fig.~\ref{fig:fig-detwinned-1}(a). Although it is likely that some anisotropy of the spectral function could be induced by the application of a sufficiently large strain when the sample temperature is close to $T_s$, in a similar spirit to the induced resistivity anisotropy observed in transport measurements under strain above $T_s$ \cite{Zhang2012b,Chu2012}, it appears that in our measurement we are not in a regime where such an induced effect can be reliably detected. 

In the twinned measurements at 45~K in the nematic phase shown in Fig.~\ref{fig:fig-detwinned-1}(c), the Fermi surface map varies only subtly from the 60 K data, showing evidence for two slightly elongated ellipses at the $\mathrm{\bar{M}}$ point. However the measurements of the detwinned sample in Fig.~\ref{fig:fig-detwinned-1}(d,e) reveal a remarkable result: \textit{in one domain, ARPES measurements only observe spectral weight on one elliptical electron pocket oriented along the longer $a_{orth}$ direction}. In Fig.~\ref{fig:fig-detwinned-1}(e) almost the whole of one elliptical electron pocket oriented along the $a_{orth}$ axis is detected at $\mathrm{\bar{M}_X}$, with no evidence for the horizontal pocket along $b_{orth}$. The sample is rotated by 90$^\circ$ in Fig.~\ref{fig:fig-detwinned-1}(d) \footnote{All data in Fig.~1 are taken with 42~eV LH photons in equivalent measurement geometries, only the detwinned sample is rotated by 90$^\circ$. Previous studies of NaFeAs did not rotate the detwinned sample, but rather changed the polarisation of the light and the orientation of the measurement to achieve similar, but not exactly equivalent measurements along $\Gamma-\mathrm{\bar{M}_Y}$ and $\Gamma-\mathrm{\bar{M}_X}$.}, where now only the left-most end of the horizontal electron pocket is observed clearly in this measurement geometry, but this is consistent with only the ellipse along $a_{orth}$ having spectral weight (although since the degree of detwinning is not 100 \% here, a small remnant intensity on the other ellipse from the minority domain population can be detected). Further measurements of detwinned samples in LV polarisation, presented in the Supplemental Material (SM) \footnote{See Supplemental Material at [URL will be inserted by publisher] for [Further ARPES spectra and discussion].}, show that the tips of the ellipses with $d_{xy}$ character also obey this ``one-ellipse" structure.

\begin{figure*}
	\centering
	\includegraphics[width=1\linewidth]{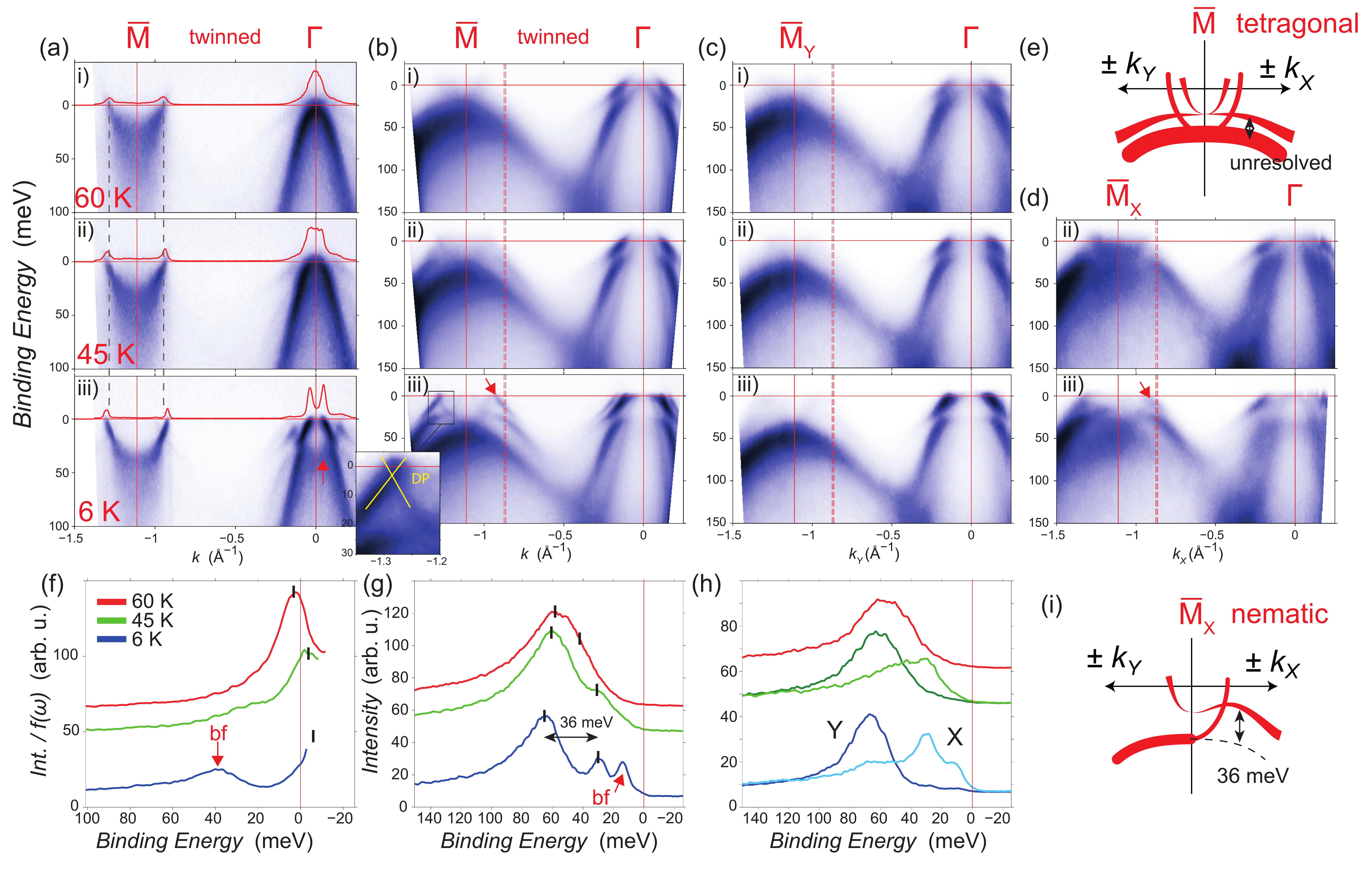}
	\caption{High-symmetry cuts in the $\Gamma$-$\mathrm{\bar{M}}$ direction, corresponding to the geometry used in Fig.~\ref{fig:fig-detwinned-1}. (a,b) Measurements of twinned samples, in LV and LH polarisations respectively, at i) 60~K in the tetragonal phase, ii) 45~K in the nematic phase and iii) at 6~K in the magnetic phase. Inset zooms in on the band crossing, indicating a Dirac point located $\sim$5~meV below $E_F$ on the $\mathrm{\Gamma-\bar{M}_X}$ line. (c,d) Measurements of detwinned samples along $\mathrm{\Gamma-\bar{M}_Y}$ and $\mathrm{\Gamma-\bar{M}_X}$ respectively, schematically represented in (e,i). (f) EDCs at the $\Gamma$ point, showing that the inner hole band crosses $E_F$. (g,h) twinned and detwinned EDCs at $k$=-0.87 \AA, corresponding to the dashed lines. X and Y refer to the EDCs obtained from $\mathrm{\Gamma-\bar{M}_X}$ and $\mathrm{\Gamma-\bar{M}_Y}$ respectively. Features labelled with red arrows and ``bf" correspond to backfolded features in the magnetic phase.}
	\label{fig:fig-detwinned-2-cuts}
\end{figure*}

In Fig.~\ref{fig:fig-detwinned-1}(f-h) the low-temperature Fermi surface maps reveal that the onset of antiferromagnetic or spin-density-wave (SDW) order has a strong impact on the observations, with backfolded features appearing in the spectra below $T_N$ \footnote{In He \textit{et al.} \cite{He2010} it was claimed that short-ranged order causes the appearance of backfolded intensity even up to $T_s$, but we do not find any evidence for this}. Hybridization with the backfolded features gaps out large sections of the Fermi surfaces, leaving a Fermi surface at both the $\Gamma$ and $\mathrm{\bar{M}}$ points containing characteristic bright ``spots". These are actually tiny quasi-2D pockets with Dirac-like dispersions \cite{Richard2010}; when the Fermi surface is reconstructed by the magnetic order, hybridisation is forbidden by symmetry along the high-symmetry directions \cite{Ran2009,Morinari2010}, enforcing a ``nodal spin-density wave" phase. The Dirac points lie on the high symmetry axes, a few meV below the Fermi level (Fig.~\ref{fig:fig-detwinned-2-cuts}(b-iii)). These tiny pockets with electron-like Dirac carriers can be associated with the large negative Hall effect and enhanced magnetoresistance observed in the magnetic phase \cite{Deng2015,Steckel2016}. Thus the Fermi surface at $\Gamma$ at low temperatures includes four tiny Fermi pockets. The tiny pockets along $a_{orth}$ have a greater separation than the pockets along $b_{orth}$, giving a characteristic rhombus-shaped constellation of bright spots in the Fermi surface maps. In addition there is a small elliptical pocket seen around the $\Gamma$ point; the three-dimensionality of this pocket will be discussed later. Due to the selection rules which strongly modulate ARPES measurements of Fe-based superconductors, not all the bright spots are always observed in any given measurement. For instance, in Fig.~\ref{fig:fig-detwinned-1}(h), the four bright spots can be clearly observed near $\mathrm{\bar{M}_X}$, while in the other sample orientation in Fig.~\ref{fig:fig-detwinned-1}(g), only the left-most of the four bright spots has significant intensity. This low temperature Fermi surface varies slightly from the determination of Zhang \textit{et al} \cite{Zhang2012b} and is consistent with the low temperature Fermi surface inferred from the quasiparticle interference measuerements of Rosenthal \textit{et al} \cite{Rosenthal2014}.

% Figure 2 discussion
In Fig.~\ref{fig:fig-detwinned-2-cuts}(a) we present the $\Gamma$-$\mathrm{\bar{M}}$ dispersions as a function of temperature. Near the $\mathrm{\bar{M}}$ point the LV polarisation highlights the end sections of the Fermi surface with $d_{xy}$ orbital character; the slight increase in $k_F$ at lower temperatures is an indication that the elliptical electron pockets elongate slightly in the nematic phase. At the $\Gamma$ point this polarisation highlights a band with predominantly  $d_{yz}$ orbital character. At 60~K the top of this band lies just a few meV below the chemical potential, but as the sample cools this band moves up and creates a 3D Fermi surface. This can also be seen from the plots of the EDCs at $\Gamma$ divided by the Fermi distribution function in Fig.~\ref{fig:fig-detwinned-2-cuts}(f). In the magnetic phase in Fig.~\ref{fig:fig-detwinned-2-cuts}(a-iii), a new feature emerges at the $\Gamma$ point, corresponding to backfolded spectral weight from the electron pocket dispersions. 

The remaining data in Fig.~\ref{fig:fig-detwinned-2-cuts}(b-d) are obtained with LH polarisation, on twinned and detwinned samples. In the previous detwinned ARPES studies \cite{Yi2012,Zhang2012b} similar data was reported, and attention was drawn to the band dispersions slightly away from the $\mathrm{\bar{M}}$ point, which we highlight by plotting the EDCs at $k$=-0.87~\AA~in Fig.~\ref{fig:fig-detwinned-2-cuts}(g). Here, at 60~K in the tetragonal phase there is a single, quite broad peak, which appears to split into two once the system enters the nematic phase; a third sharp peak from the backfolded outer hole band appears only the magnetic phase. Previously, this apparent splitting, reaching 36 meV at low temperatures, was considered to be a direct consequence of orbital order: it was suggested that in the nematic phase the $d_{xz}$ and $d_{yz}$ orbitals lose their degeneracy, thus the band dispersions towards $\mathrm{\bar{M}_Y}$ and $\mathrm{\bar{M}_X}$ with $d_{xz/yz}$ orbital character become non-equivalent, manifesting as a splitting in measurements of twinned samples. Despite the additional complications of backfolded features in the magnetic phase, the underlying separate band dispersions can still be seen. Since very similar effects were claimed in BaFe$_2$As$_2$ \cite{Yi2011} and a study of detwinned FeSe \cite{Shimojima2014}, this ``orbital splitting" as determined by (detwinned) ARPES measurements has become widely regarded as a hallmark of nematic order. 

However, there are some shortcomings of this scenario. For instance, although a large 30-40 meV energy scale has been claimed for this splitting \cite{Yi2012,Zhang2012b}, no such large energy scale is observed at the $\Gamma$ point, and it seems large compared with $T_s$ = 54~K $\approx$ 5~meV. Additionally, one has to assume that the second band with $d_{xy}$ character which is expected to disperse from  $\Gamma$-$\mathrm{\bar{M}}$ does not contribute to the measurement, i.e. that ARPES does not observe all the bands. Yet we have shown that at 60~K the full Fermi surface consisting of two crossed ellipses may be observed, including sections with $d_{xy}$ character, and recent measurements of FeSe have shown that this second band clearly does contribute brightly to the dispersions in this geometry in the tetragonal phase \cite{Watson2016,Fedorov2016}. Instead, we suggest that the distinct dispersions seen in NaFeAs along $\mathrm{\Gamma-M_X}$ and $\mathrm{\Gamma-M_Y}$ in Fig.~\ref{fig:fig-detwinned-2-cuts}(c,d) are actually a natural extension of the ``one-ellipse" structure in the nematic phase. We have already shown that only the elliptical Fermi surface oriented along the $a_{orth}$ pocket is observed by ARPES, and moreover this extends to the occupied states, where \textit{only half of the expected dispersions near the ${\bar{M}}$ point have spectral weight in the nematic phase} - the dispersions which are necessary to constitute one elliptical Fermi surface. 

\begin{figure}
	\centering
	\includegraphics[width=\linewidth]{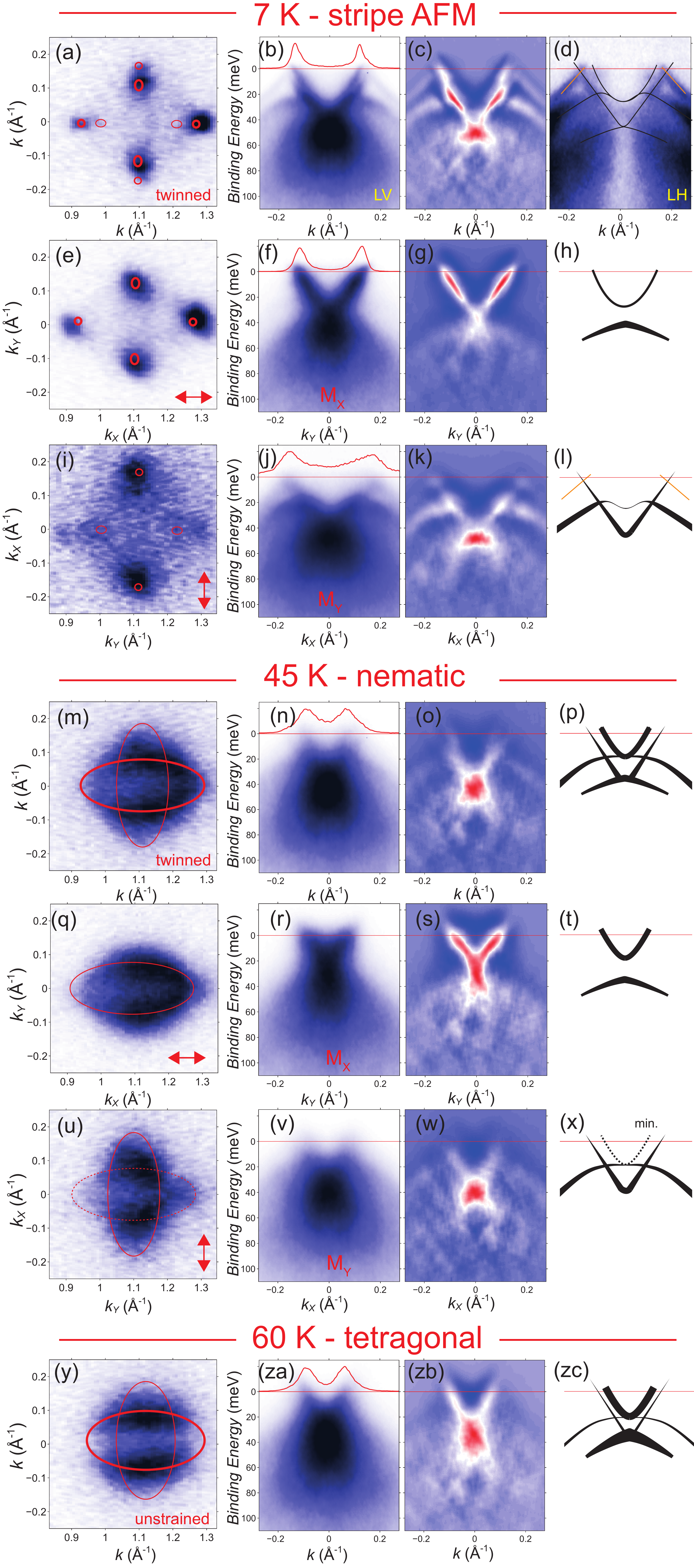}
	\caption{Twinned and detwinned measurements at the M point, using 47 eV photons. a) Low temperature Fermi surface map of a twinned sample, b) high-symmetry cut in LV polarisation, c) curvature plot of the data from (b). d) Schematic bands, here overlaid over measurements obtained in LH polarisation. Backfolded features are marked in orange color. (e-i) As (a-d), but for the two different sample orientations of a detwinned sample. The direction of tensile strain is indicated by the red double-headed arrows. (m-x) Equivalent measurements at 45~K in the nematic phase and (y-zc) at 60~K in the tetragonal phase.}
	\label{fig:fig-detwinned-3-m}
\end{figure}

Let us develop this a little further. In the tetragonal phase, one would expect to see two bands with $d_{xz}$ and $d_{xy}$ orbital character dispersing from $\Gamma$-$\mathrm{\bar{M}}$ - the hole-like dispersions in Fig.~\ref{fig:fig-detwinned-2-cuts}(e). We suggest that both bands contribute to the measurement at 60~K in the tetragonal phase in Fig.~\ref{fig:fig-detwinned-2-cuts}(b-i), but they are unresolved since their separation is less than the energy resolution; the complex lineshape  of the EDC observed in Fig.~\ref{fig:fig-detwinned-2-cuts}(g) with a 'shoulder' at low binding energy hints that this is likely to be the case. Then, in the nematic phase, the ``one-ellipse" structure kicks in, and  correspondingly the dispersion at lower binding energy is observed along $\Gamma$-$\mathrm{\bar{M_X}}$ and the deeper dispersion is observed along $\Gamma$-$\mathrm{\bar{M_Y}}$. We note that this is a completely distinct effect from any orbital splitting. On top of this, the bands \textit{do} have an increased separation (not a splitting) in the nematic phase, related to the elongation of the Fermi surface and an indication that there exists some form of nematic or orbital ordering on a $\sim$10-15~meV scale (not 30-40 meV), but we emphasize that the \textit{primary result of the measurements is the ``one-ellipse'' structure}, with correspondingly one dispersion observed from $\mathrm{\bar{M}}$ towards $\Gamma$ in the occupied states. This is schematically shown in Fig.~\ref{fig:fig-detwinned-2-cuts}(i). While non-intuitive, these observations exactly mirror the recent result of high-resolution detwinned ARPES measurements of nematic FeSe \cite{Watson2017d_arxiv}. We therefore suggest that the one-ellipse structure of the observed spectral weight may be generic to all Fe-based superconductors which enter the nematic phase.

In Fig.~\ref{fig:fig-detwinned-3-m} we present further ARPES spectra, now focusing only on the electron pockets at the M point. The low temperature Fermi surface consists of four tiny pockets appearing as bright spots, since the electronic structure is backfolded and reconstructed. However since hybridisation does not occur on the high-symmetry directions, the high-symmetry measurements in Fig.~\ref{fig:fig-detwinned-3-m}(b-d) mainly probes the spectral weight from the underlying electron pocket dispersions, with additional features from backfolded hole bands. For the twinned sample in Fig.~\ref{fig:fig-detwinned-3-m}(b-d), the curvature analysis helps to reveal four (non-backfolded) band dispersions including two bands extending up to $E_F$, resembling ARPES spectra in overdoped NaFeAs \cite{Charnukha2016}. However the detwinned measurements in Fig.~\ref{fig:fig-detwinned-3-m}(e-l) reveal that actually only two dispersions are observed in one domain. Thus despite the complication of magnetic backfolding and reconstruction of the Fermi surface, the high-symmetry cuts at M again reveal that the underlying band dispersions correspond to only one (now reconstructed) elliptical pocket along $a_{orth}$. 

At 45~K, the features are significantly broader, but the curvature analysis helps to show that four band dispersions contribute in the twinned sample, shown in Fig.~\ref{fig:fig-detwinned-3-m}(n-p). These four band dispersions correspond to the two crossed ellipses which we draw in Fig.~\ref{fig:fig-detwinned-3-m}(m), however at the Fermi level the intensity of the outer dispersion becomes quite low, meaning that the Fermi surface map is dominated by the horizontal ellipse. In the detwinned geometry in Fig.~\ref{fig:fig-detwinned-3-m}(q), only this horizontal ellipse oriented along the $a_{orth}$ direction is observed, and correspondingly two dispersions are seen in the high-statistics cut in Fig.~\ref{fig:fig-detwinned-3-m}(r-t). Upon rotating the sample in Fig.~\ref{fig:fig-detwinned-3-m}(u-x) the spectral weight is predominantly on the vertical ellipse - although here there is a contamination of the signal from the minority orthorhombic domain near the Fermi level due to the intrinsically lower intensity of the outer band dispersion from the majority domain. Thus also in this measurement geometry, the ``one-ellipse" observation of the electron pockets is the defining feature of the nematic phase.

\begin{figure*}
	\centering
	\includegraphics[width=0.6\linewidth]{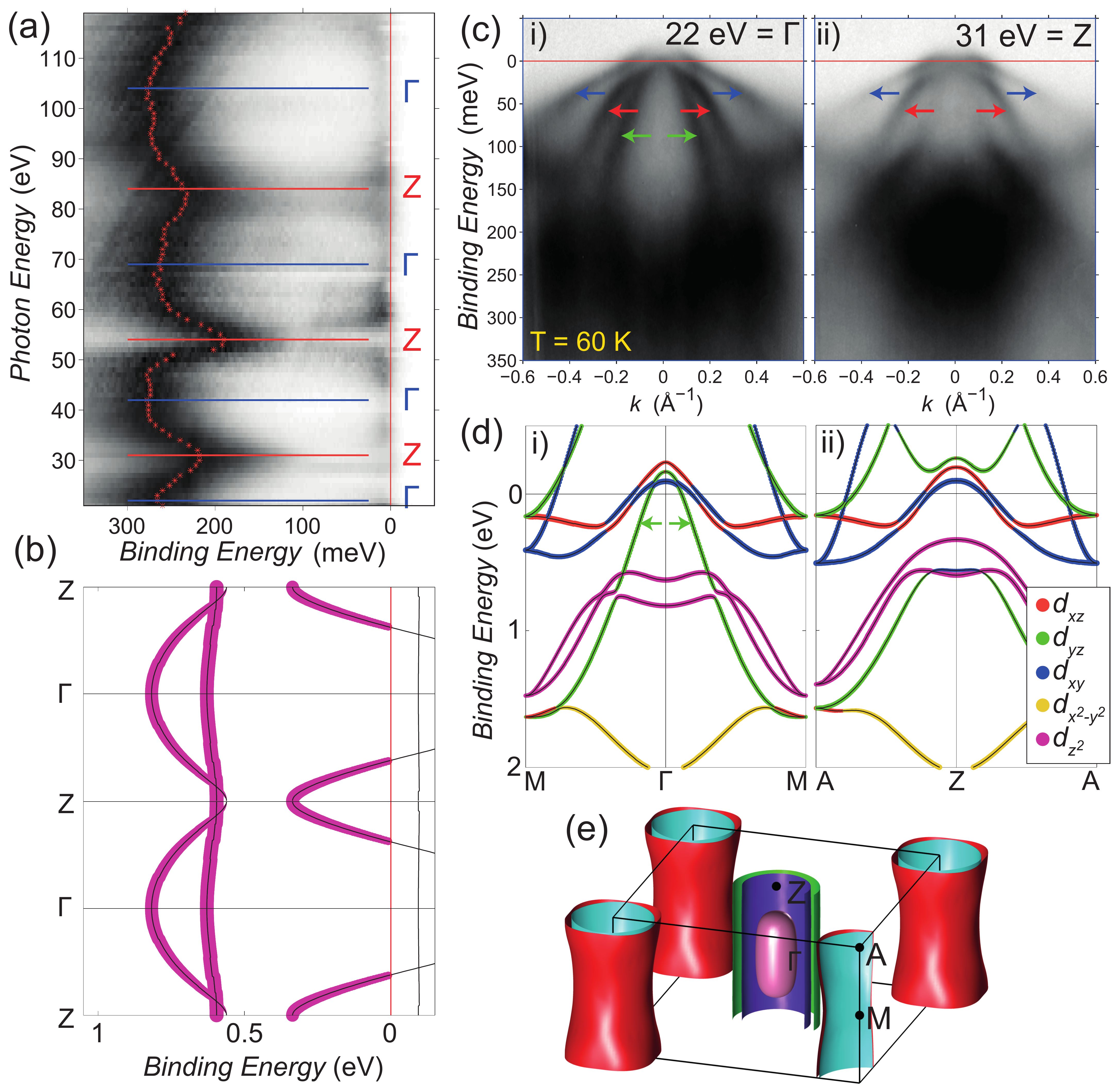}
	\caption[$k_z$]{\textbf{Analysis of $k_z$ variation of NaFeAs.} (a) Plot of the EDC at normal emission as a function of photon energy. Data were obtained in LH polarisation, at 60~K in the tetragonal phase.  Red stars indicate the location of maximum intensity after data smoothing. (b) DFT calculation along the $\Gamma$-Z-$\Gamma$ direction, highlighting the $d_{z^2}$ orbital weight. (c) High-symmetry cuts at 22 eV ($\Gamma$) and 31 eV (Z). The presence of the $d_{yz}$ band at the $\Gamma$ point only is indicated by the green arrows. (d) DFT band structures for the M-$\Gamma$-M and A-Z-A paths. (e) Fermi surface of NaFeAs according to DFT. While the experimental electronic structure of NaFeAs varies substantially, DFT correctly captures the formation of a 3D hole pocket centred at $\Gamma$.}
	\label{fig:fig-kz}
\end{figure*}

At 60~K, in the tetragonal phase, the spectra in Fig.~\ref{fig:fig-detwinned-3-m}(y-zc) become rather broad, making it hard to distinguish separate band dispersions (SM). However we believe that all four expected band dispersions contribute to this spectrum in the tetragonal phase, even if the EDCs are too broad to directly show the separation of bands. This scenario would be consistent with the observations of FeSe in the tetragonal phase \cite{Watson2016,Fedorov2016}.

\subsection{Backfolding of spectral weight by ($\bf \pi$,0,$\bf \pi$)}

\begin{figure*}
	\centering
	\includegraphics[width=0.95\linewidth]{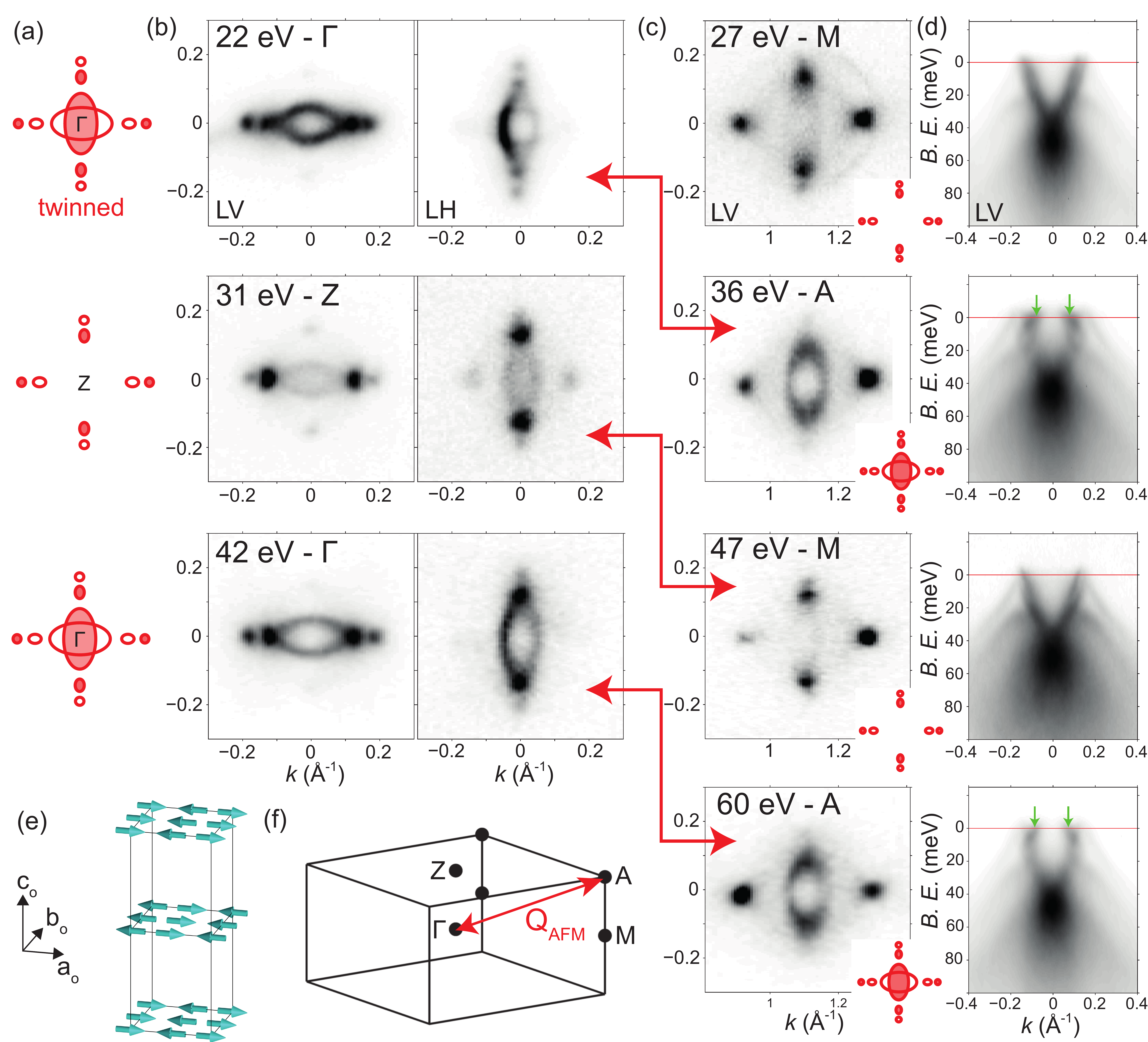}
	\caption{(a) Schematic slices of the low temperature Fermi surface of NaFeAs in the magnetic phase. For comparison with ARPES measurements of twinned samples we show two domains, with one domain being shaded. (b) Corresponding Fermi surface maps at the $\Gamma$ and Z points of the 3D Brillouin zone, in LV and LH polarisation. (c) Fermi surface maps at the M and A points, and (d) corresponding high-symmetry cuts. The clear qualitative difference as a function of $k_z$ is the presence of the elliptical Fermi surface at the A point, arising from the backfolding of the hole pocket at $\Gamma$. (e) Magnetic configuration of NaFeAs, based on Li \textit{et al.} \cite{Li2009} (f) The wavevector $Q_{AFM}$ of the magnetic backfolding in the 3D Brillouin zone.}
	\label{fig:fig-backfolding}
\end{figure*}

We now change focus and discuss an issue which has been previously overlooked in ARPES measurements of NaFeAs: the $k_z$ component of the backfolding. Firstly, we experimentally determine the $k_z$ dependence in the tetragonal phase, by plotting the variation of the EDC at normal emission as a function of photon energy at 60~K in Fig.~\ref{fig:fig-kz}(a). The spectrum is dominated by a broad peak at $\sim$ 200-250 meV, corresponding to the $d_{z^2}$ orbital weight. In the corresponding DFT calculation in Fig.~\ref{fig:fig-kz}(b) it is predicted that there should be a pair of bands, but due to a significant imaginary component of the self energy at these binding energies \cite{Nekrasov2015} only one broad peak is observed experimentally. As is well-known, the variation of the photon energy in ARPES measurements allows one to effectively measure at different $k_z$, and the periodic variation of the $d_{z^2}$ feature is a convenient feature from which we infer appropriate photon energies for the $\Gamma$ and Z points, broadly consistent with Refs.~\cite{He2011,Thirupathaiah2012,Liu2012,Ge2013}. 

DFT calculations for NaFeAs also predict another band which disperses rather strongly along $k_z$, passing through the Fermi level along the $\Gamma$-Z line, as shown in Fig.~\ref{fig:fig-kz}(b). This band, with largely Fe $d_{z^2}$ and some As $p_z$ character has been previously observed in LiFeAs \cite{Borisenko2015} and is detected in our data, being especially clear around 50-60 eV in Fig.~\ref{fig:fig-kz}(a). The dispersion of this band is not universal in Fe-based superconductors, for instance the band remains above the Fermi level in FeSe \cite{Watson2015}, but the existence of a similar band in Fe(Te$_{0.55}$Se$_{0.45}$) is proposed to be responsible for the band inversion and related topological character and surface states in that system \cite{Zhang2017_arxiv,Wang2015top}. The presence of this $k_z$-dispersing band in the DFT calculation effectively truncates the innermost hole-like cylindrical Fermi surface, giving rise to a 3D hole pocket as shown in Fig.~\ref{fig:fig-kz}(e). The expected dispersions along the M-$\Gamma$-M and A-Z-A cuts are also qualitatively different: as shown in Fig.~\ref{fig:fig-kz}(d) there are three hole-like bands around the Fermi level at $\Gamma$, but at the Z point there are only two, with the $d_{yz}$ band being absent. This qualitative difference is also present in the comparison of ARPES spectra at 22 eV and 31 eV, identified as $\Gamma$ and Z respectively; the $d_{yz}$ band indicated by the green arrows in Fig.~\ref{fig:fig-kz}(c-i) is absent Fig.~\ref{fig:fig-kz}(c-ii).

The existence of an additional band at the $\Gamma$ point which is not present at the Z point has interesting implications in the magnetically reconstructed phase. 
At the $\Gamma$ point, this band essentially survives into the magnetic phase, becoming somewhat elongated. Evidence that this is also three-dimensional pocket can be seen in  Fig.~\ref{fig:fig-backfolding}(b); at 22 eV and 42 eV which both correspond to $\Gamma$ points the elliptical pocket is present, while at 31 eV corresponding to the Z point, only a faint shadow is observed. The faint shadow at the Z point arises from the uncertainty in $k_z$ in photoemission as well as possible higher-order backfolding effects (further discussed in SM). This creates a clear distinction between the $\Gamma$ and Z planes; both contain very tiny electron-like Fermi surfaces, but additionally there is significant spectral weight on the elliptical hole-like pocket derived from the inner band at $\Gamma$, but it is absent at the Z point, shown schematically in Fig.~\ref{fig:fig-backfolding}(a).

In Fig.~\ref{fig:fig-backfolding}(c) we present low temperature Fermi surface maps at the M and A points. In addition to the tiny spot-like Fermi surfaces, an elliptical hole-like band is present at each A point, but not at the M points \footnote{An additional rather faint outer ring of intensity is detected in the 27 eV data in Fig.~\ref{fig:fig-backfolding}(c). The origin of this feature is not clear to us but we do not believe that it is likely to be a bulk quasiparticle band.}. This can also be seen in the high-symmetry cuts presented in Fig.~\ref{fig:fig-backfolding}(d), where the green arrows indicate the backfolded hole band which is only present in the magnetic phase at the A point. This elliptical pocket which appears only at A is therefore a backfolded copy of the 3D elliptical pocket which is only present at $\Gamma$. This proves that the A point maps onto the $\Gamma$ point in the reconstructed Brillouin zone, i.e. that the magnetic reconstruction has the wavevector ($\pi$,0,$\pi$), as shown schematically in Fig.~\ref{fig:fig-backfolding}(f). 

\begin{figure}[t!]
	\centering
	\includegraphics[width=\linewidth]{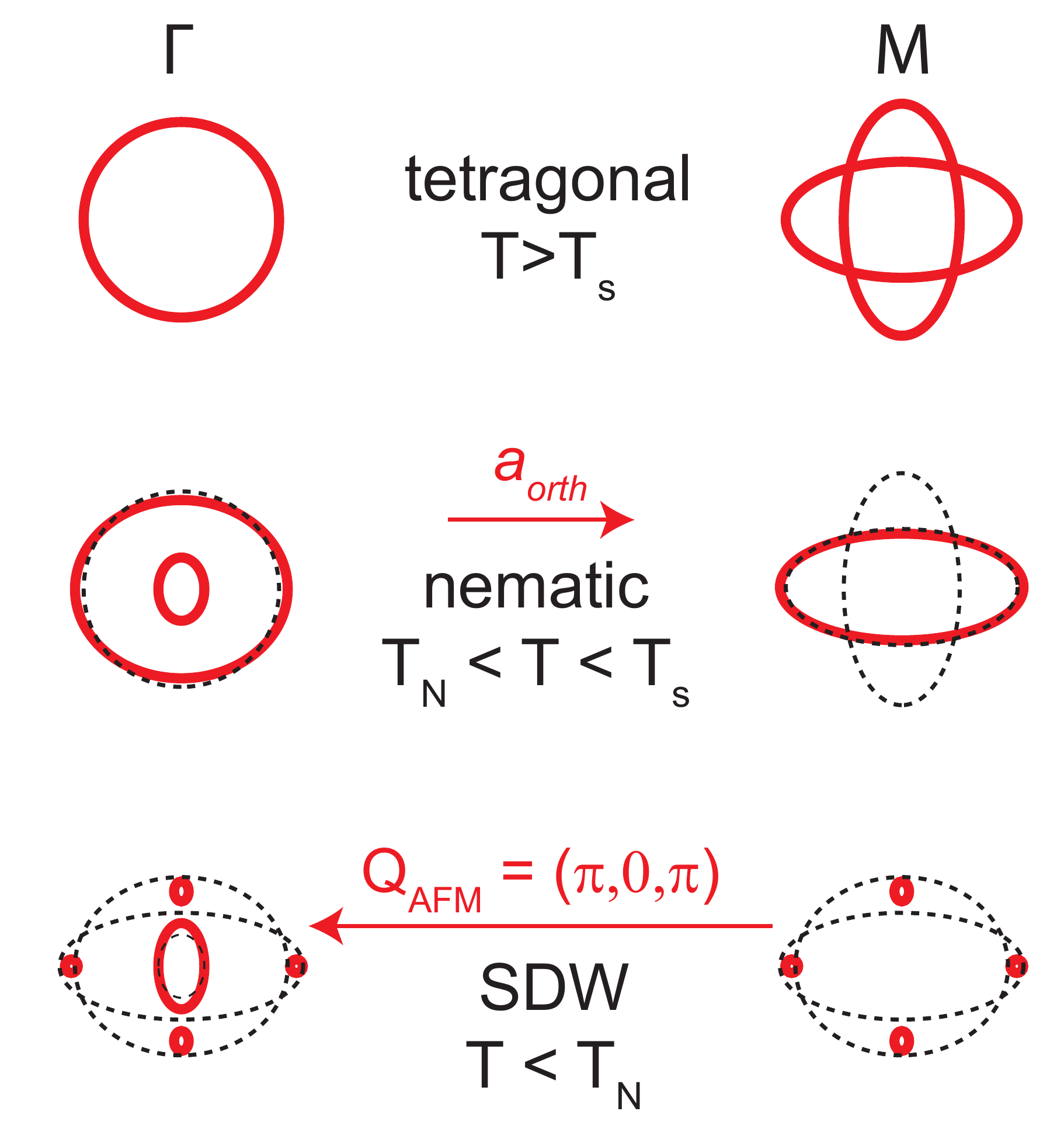}
	\caption{Top: Schematic Fermi surfaces of NaFeAs at the $\Gamma$ and $\mathrm{M}$ points, as measured by ARPES in the tetragonal phase of FeSe. Middle:	Schematic for the nematic phase. Black dotted lines represent the Fermi surfaces from the tetragonal phase, showing the absence of the pocket along $b_{orth}$ and more subtle Fermi surface elongations. Bottom: The reconstructed electronic structure in the SDW phase. Black dashed lines are overlapped electron and hole bands from the nematic phase. Note that in the SDW phase, the Fermi surfaces at $\Gamma$ and M are not equivalent, since the $\Gamma$ point actually maps to the A point.}
	\label{fig:fig-schematicsummary}
\end{figure}

Our observations are consistent with the determination of the magnetic structure of NaFeAs by neutron scattering measurements. In common with other parent compounds of Fe-based superconductors, the in-plane magnetic configuration of NaFeAs is stripe-like, antiferromagnetic along the longer $a_{orth}$ axis and ferromagnetic along $b_{orth}$ \cite{Dai2015review}. However neutron scattering measurements by Li \textit{et al} \cite{Li2009} established that there is also a $c$-axis component of the ordering; The antiferromagnetic phase of NaFeAs has ``C-type" magnetic configuration as shown in Fig.~\ref{fig:fig-backfolding}(e), with a phase shift of the magnetic stripes between the layers. Therefore in reciprocal space, the antiferromagnetic ordering vector $Q_{AFM}$ can be written as ($\pi$,0,$\pi$) in the 1-Fe unit cell notation. This manifests in ARPES measurements in the backfolded features described above.

\section{Discussion}

We summarise our ARPES results in Fig.~\ref{fig:fig-schematicsummary}. Above $T_s$, we have a simple Fermi surface with a single hole pocket, and two crossed ellipses are detected at the M point. Upon entering the nematic phase, the most dramatic effect is that now only the electron pocket oriented along the longer $a_{orth}$ is observed. In addition, the Fermi surface undergoes some subtle distortions, with an elongation of the electron pocket and the incipient inner hole band now crossing the Fermi level. Finally in the SDW phase, this one electron pocket is backfolded along the antiferromagnetic $a_{orth}$ direction onto the hole bands, hybridising everywhere except along the high-symmetry directions, leading to the characteristic tiny spot-like Fermi surfaces with Dirac-like dispersions. However importantly, the backfolding has a $c$-axis component; as the wavevector of the backfolding is ($\pi$,0,$\pi$) in the 2-Fe unit cell, it is the $\mathrm{A}$ point which maps to the $\Gamma$ point, not the $\mathrm{M}$ point. 

Our results give us confidence that we can probe the true bulk 3D electronic structure of NaFeAs. Due to the surface-sensitivity of ARPES, one could worry about whether the backfolded bands could instead be arising from photoelectrons which are scattered by a \textit{surface} reciprocal lattice vector due to the superstructure in the magnetic phase. However the observation of the $k_z$ component of the backfolding in the magnetic phase indicates that ARPES is probing the bulk spectral function. 

It is worthwhile to compare our detwinned ARPES results in NaFeAs with recent high-resolution detwinned measurements of FeSe \cite{Watson2017d_arxiv}. The observed elliptical electron pocket elongates in the nematic phase in both cases, although this is substantially more pronounced in the case of FeSe. On the other hand, in FeSe the outer elliptical hole band in FeSe is longer in the $b_{orth}$ direction \cite{Watson2017d_arxiv}, whereas in NaFeAs the outer hole pocket is longer along the $a_{orth}$ axis, suggesting a possible lack of universality in the form of the orbital ordering. However in any case, we would argue the details of the magnitude and momentum-dependence of the orbital/nematic ordering are not of primary importance, but instead we propose that the ``one ellipse" structure of the spectral weight which has now been observed in both materials is the defining feature of the electronic structure of the nematic phase. 

The dramatic anisotropy associated with the ``one ellipse" structure of the spectral weight in the nematic phases of NaFeAs and FeSe cannot be accounted for by orbital ordering alone. Instead we suggest that one must look for an explanation involving a coupling of the itinerant electrons to strongly anisotropic magnetic fluctuations in the nematic phase. In this view, the structural transition is the temperature at which the stripe-like magnetic fluctuations are sufficiently strong to break the Ising-nematic $\mathbb{Z}_2$ symmetry, but their correlation length remains finite \cite{Wright2012,Ma2011,Rosenthal2014}; possibly the inter-plane coherence is not yet established. When the system has determined its nematic orientation, this also dictates which electron pocket preserves the spectral weight. Once this choice is made, the characteristically anisotropic properties of the nematic phase follow. The one-ellipse structure would naturally lead to a marked in-plane resistivity anisotropy \cite{Zhang2012b,Deng2015}. Moreover the low-energy spin susceptibility will be much larger along the $a_{orth}$ direction; the recent observation of neutron spin-resonances at ($\pi$,0,$\pi$) but no intensity at (0,$\pi$,$\pi$) in detwinned inelastic neutron scattering measurements of Na(Fe$_{0.985}$Co$_{0.015}$)As \cite{Wang2017} could be compatible with this picture. If the system remains non-magnetic but becomes superconducting, the absence of states to pair with on the electron pocket along $b_{orth}$ would naturally lead to a substantially anisotropic twofold symmetric gap structure within the context of a spin-fluctuation pairing mechanism, which could explain recent results in FeSe \cite{Sprau2017}.  At present we do not have a detailed understanding of the microscopic mechanism or symmetry argument which links the suppression of spectral weight on the electron pocket along $b_{orth}$ with the magnetically-driven nematic state, but we would encourage theoretical studies of this unexpected - but defining - feature of the nematic phase.

\section{Conclusion}

In conclusion, we have revisited the experimental determination of the electronic structure of NaFeAs with high-resolution twinned and detwinned ARPES measurements. The defining feature of the nematic phase is the observation of spectral weight only on the elliptical electron pocket oriented along the longer $a_{orth}$ axis of the orthorhombic structure. In the SDW phase, this one electron pocket is backfolded by the antiferromagnetic wavevector of ($\pi$,0,$\pi$), leaving an electronic structure with four tiny 2D pockets on the high symmetry axes and a three-dimensional hole-like band around the $\Gamma$ point. Our measurements clarify how the SDW phase of NaFeAs emerges, and indicate that the nematic phase is driven by magnetic, not orbital, degrees of freedom. 

\begin{acknowledgments}
	\section{acknowledgments}
	We thank S.~V.~Borisenko and P.~D.~C.~King for useful discussions. We thank P.~Dudin and D.~Daisenberger for technical assistance. We thank Diamond Light Source for access to Beamline I05 (Proposals No. SI15074, NT18218, NT15663) that contributed to the results presented here. IM acknowledges support by the RSF grant No.16-42-01100.
\end{acknowledgments}

%\bibliography{NaFeAs_bib}  

\begin{thebibliography}{57}%
	\makeatletter
	\providecommand \@ifxundefined [1]{%
		\@ifx{#1\undefined}
	}%
	\providecommand \@ifnum [1]{%
		\ifnum #1\expandafter \@firstoftwo
		\else \expandafter \@secondoftwo
		\fi
	}%
	\providecommand \@ifx [1]{%
		\ifx #1\expandafter \@firstoftwo
		\else \expandafter \@secondoftwo
		\fi
	}%
	\providecommand \natexlab [1]{#1}%
	\providecommand \enquote  [1]{``#1''}%
	\providecommand \bibnamefont  [1]{#1}%
	\providecommand \bibfnamefont [1]{#1}%
	\providecommand \citenamefont [1]{#1}%
	\providecommand \href@noop [0]{\@secondoftwo}%
	\providecommand \href [0]{\begingroup \@sanitize@url \@href}%
	\providecommand \@href[1]{\@@startlink{#1}\@@href}%
	\providecommand \@@href[1]{\endgroup#1\@@endlink}%
	\providecommand \@sanitize@url [0]{\catcode `\\12\catcode `\$12\catcode
		`\&12\catcode `\#12\catcode `\^12\catcode `\_12\catcode `\%12\relax}%
	\providecommand \@@startlink[1]{}%
	\providecommand \@@endlink[0]{}%
	\providecommand \url  [0]{\begingroup\@sanitize@url \@url }%
	\providecommand \@url [1]{\endgroup\@href {#1}{\urlprefix }}%
	\providecommand \urlprefix  [0]{URL }%
	\providecommand \Eprint [0]{\href }%
	\providecommand \doibase [0]{http://dx.doi.org/}%
	\providecommand \selectlanguage [0]{\@gobble}%
	\providecommand \bibinfo  [0]{\@secondoftwo}%
	\providecommand \bibfield  [0]{\@secondoftwo}%
	\providecommand \translation [1]{[#1]}%
	\providecommand \BibitemOpen [0]{}%
	\providecommand \bibitemStop [0]{}%
	\providecommand \bibitemNoStop [0]{.\EOS\space}%
	\providecommand \EOS [0]{\spacefactor3000\relax}%
	\providecommand \BibitemShut  [1]{\csname bibitem#1\endcsname}%
	\let\auto@bib@innerbib\@empty
	%</preamble>
	\bibitem [{\citenamefont {Mazin}\ \emph {et~al.}(2008)\citenamefont {Mazin},
		\citenamefont {Singh}, \citenamefont {Johannes},\ and\ \citenamefont
		{Du}}]{Mazin2008}%
	\BibitemOpen
	\bibfield  {author} {\bibinfo {author} {\bibfnamefont {I.~I.}\ \bibnamefont
			{Mazin}}, \bibinfo {author} {\bibfnamefont {D.~J.}\ \bibnamefont {Singh}},
		\bibinfo {author} {\bibfnamefont {M.~D.}\ \bibnamefont {Johannes}}, \ and\
		\bibinfo {author} {\bibfnamefont {M.~H.}\ \bibnamefont {Du}},\ }\bibfield
	{title} {\enquote {\bibinfo {title} {{Unconventional Superconductivity with a
					Sign Reversal in the Order Parameter of LaFeAsO$_{1-x}$F$_x$}},}\ }\href
	{\doibase 10.1103/PhysRevLett.101.057003} {\bibfield  {journal} {\bibinfo
			{journal} {Phys. Rev. Lett.}\ }\textbf {\bibinfo {volume} {101}},\ \bibinfo
		{pages} {057003} (\bibinfo {year} {2008})}\BibitemShut {NoStop}%
	\bibitem [{\citenamefont {Hirschfeld}\ \emph {et~al.}(2011)\citenamefont
		{Hirschfeld}, \citenamefont {Korshunov},\ and\ \citenamefont
		{Mazin}}]{Hirschfeld2011}%
	\BibitemOpen
	\bibfield  {author} {\bibinfo {author} {\bibfnamefont {P~J}\ \bibnamefont
			{Hirschfeld}}, \bibinfo {author} {\bibfnamefont {M~M}\ \bibnamefont
			{Korshunov}}, \ and\ \bibinfo {author} {\bibfnamefont {I~I}\ \bibnamefont
			{Mazin}},\ }\bibfield  {title} {\enquote {\bibinfo {title} {{Gap symmetry and
					structure of Fe-based superconductors}},}\ }\href
	{http://stacks.iop.org/0034-4885/74/i=12/a=124508} {\bibfield  {journal}
		{\bibinfo  {journal} {Reports on Progress in Physics}\ }\textbf {\bibinfo
			{volume} {74}},\ \bibinfo {pages} {124508} (\bibinfo {year}
		{2011})}\BibitemShut {NoStop}%
	\bibitem [{\citenamefont {Li}\ \emph {et~al.}(2009)\citenamefont {Li},
		\citenamefont {de~la Cruz}, \citenamefont {Huang}, \citenamefont {Chen},
		\citenamefont {Xia}, \citenamefont {Luo}, \citenamefont {Wang},\ and\
		\citenamefont {Dai}}]{Li2009}%
	\BibitemOpen
	\bibfield  {author} {\bibinfo {author} {\bibfnamefont {Shiliang}\
			\bibnamefont {Li}}, \bibinfo {author} {\bibfnamefont {Clarina}\ \bibnamefont
			{de~la Cruz}}, \bibinfo {author} {\bibfnamefont {Q.}~\bibnamefont {Huang}},
		\bibinfo {author} {\bibfnamefont {G.~F.}\ \bibnamefont {Chen}}, \bibinfo
		{author} {\bibfnamefont {T.-L.}\ \bibnamefont {Xia}}, \bibinfo {author}
		{\bibfnamefont {J.~L.}\ \bibnamefont {Luo}}, \bibinfo {author} {\bibfnamefont
			{N.~L.}\ \bibnamefont {Wang}}, \ and\ \bibinfo {author} {\bibfnamefont
			{Pengcheng}\ \bibnamefont {Dai}},\ }\bibfield  {title} {\enquote {\bibinfo
			{title} {{Structural and magnetic phase transitions in
					${\text{Na}}_{1\ensuremath{-}\ensuremath{\delta}}\text{FeAs}$}},}\ }\href
	{\doibase 10.1103/PhysRevB.80.020504} {\bibfield  {journal} {\bibinfo
			{journal} {Phys. Rev. B}\ }\textbf {\bibinfo {volume} {80}},\ \bibinfo
		{pages} {020504} (\bibinfo {year} {2009})}\BibitemShut {NoStop}%
	\bibitem [{\citenamefont {Presniakov}\ \emph {et~al.}(2013)\citenamefont
		{Presniakov}, \citenamefont {Morozov}, \citenamefont {Sobolev}, \citenamefont
		{Roslova}, \citenamefont {Boltalin}, \citenamefont {Son}, \citenamefont
		{Volkova}, \citenamefont {Vasiliev}, \citenamefont {Wurmehl},\ and\
		\citenamefont {Büchner}}]{Presniakov2013}%
	\BibitemOpen
	\bibfield  {author} {\bibinfo {author} {\bibfnamefont {I}~\bibnamefont
			{Presniakov}}, \bibinfo {author} {\bibfnamefont {I}~\bibnamefont {Morozov}},
		\bibinfo {author} {\bibfnamefont {A}~\bibnamefont {Sobolev}}, \bibinfo
		{author} {\bibfnamefont {M}~\bibnamefont {Roslova}}, \bibinfo {author}
		{\bibfnamefont {A}~\bibnamefont {Boltalin}}, \bibinfo {author} {\bibfnamefont
			{V}~\bibnamefont {Son}}, \bibinfo {author} {\bibfnamefont {O}~\bibnamefont
			{Volkova}}, \bibinfo {author} {\bibfnamefont {A}~\bibnamefont {Vasiliev}},
		\bibinfo {author} {\bibfnamefont {S}~\bibnamefont {Wurmehl}}, \ and\ \bibinfo
		{author} {\bibfnamefont {B}~\bibnamefont {Büchner}},\ }\bibfield  {title}
	{\enquote {\bibinfo {title} {Local structure and hyperfine interactions of 57
				fe in nafeas studied by mössbauer spectroscopy},}\ }\href
	{http://stacks.iop.org/0953-8984/25/i=34/a=346003} {\bibfield  {journal}
		{\bibinfo  {journal} {Journal of Physics: Condensed Matter}\ }\textbf
		{\bibinfo {volume} {25}},\ \bibinfo {pages} {346003} (\bibinfo {year}
		{2013})}\BibitemShut {NoStop}%
	\bibitem [{\citenamefont {Kontani}\ and\ \citenamefont
		{Onari}(2010)}]{Kontani2010}%
	\BibitemOpen
	\bibfield  {author} {\bibinfo {author} {\bibfnamefont {Hiroshi}\ \bibnamefont
			{Kontani}}\ and\ \bibinfo {author} {\bibfnamefont {Seiichiro}\ \bibnamefont
			{Onari}},\ }\bibfield  {title} {\enquote {\bibinfo {title}
			{{Orbital-Fluctuation-Mediated Superconductivity in Iron Pnictides: Analysis
					of the Five-Orbital Hubbard-Holstein Model}},}\ }\href {\doibase
		10.1103/PhysRevLett.104.157001} {\bibfield  {journal} {\bibinfo  {journal}
			{Phys. Rev. Lett.}\ }\textbf {\bibinfo {volume} {104}},\ \bibinfo {pages}
		{157001} (\bibinfo {year} {2010})}\BibitemShut {NoStop}%
	\bibitem [{\citenamefont {Yi}\ \emph {et~al.}(2011)\citenamefont {Yi},
		\citenamefont {Lu}, \citenamefont {Chu}, \citenamefont {Analytis},
		\citenamefont {Sorini}, \citenamefont {Kemper}, \citenamefont {Moritz},
		\citenamefont {Mo}, \citenamefont {Moore}, \citenamefont {Hashimoto},
		\citenamefont {Lee}, \citenamefont {Hussain}, \citenamefont {Devereaux},
		\citenamefont {Fisher},\ and\ \citenamefont {Shen}}]{Yi2011}%
	\BibitemOpen
	\bibfield  {author} {\bibinfo {author} {\bibfnamefont {Ming}\ \bibnamefont
			{Yi}}, \bibinfo {author} {\bibfnamefont {Donghui}\ \bibnamefont {Lu}},
		\bibinfo {author} {\bibfnamefont {Jiun-Haw}\ \bibnamefont {Chu}}, \bibinfo
		{author} {\bibfnamefont {James~G.}\ \bibnamefont {Analytis}}, \bibinfo
		{author} {\bibfnamefont {Adam~P.}\ \bibnamefont {Sorini}}, \bibinfo {author}
		{\bibfnamefont {Alexander~F.}\ \bibnamefont {Kemper}}, \bibinfo {author}
		{\bibfnamefont {Brian}\ \bibnamefont {Moritz}}, \bibinfo {author}
		{\bibfnamefont {Sung-Kwan}\ \bibnamefont {Mo}}, \bibinfo {author}
		{\bibfnamefont {Rob~G.}\ \bibnamefont {Moore}}, \bibinfo {author}
		{\bibfnamefont {Makoto}\ \bibnamefont {Hashimoto}}, \bibinfo {author}
		{\bibfnamefont {Wei-Sheng}\ \bibnamefont {Lee}}, \bibinfo {author}
		{\bibfnamefont {Zahid}\ \bibnamefont {Hussain}}, \bibinfo {author}
		{\bibfnamefont {Thomas~P.}\ \bibnamefont {Devereaux}}, \bibinfo {author}
		{\bibfnamefont {Ian~R.}\ \bibnamefont {Fisher}}, \ and\ \bibinfo {author}
		{\bibfnamefont {Zhi-Xun}\ \bibnamefont {Shen}},\ }\bibfield  {title}
	{\enquote {\bibinfo {title} {Symmetry-breaking orbital anisotropy observed
				for detwinned ba(fe1-xcox)2as2 above the spin density wave transition},}\
	}\href {\doibase 10.1073/pnas.1015572108} {\bibfield  {journal} {\bibinfo
			{journal} {Proceedings of the National Academy of Sciences}\ }\textbf
		{\bibinfo {volume} {108}},\ \bibinfo {pages} {6878--6883} (\bibinfo {year}
		{2011})},\ \Eprint
	{http://arxiv.org/abs/http://www.pnas.org/content/108/17/6878.full.pdf}
	{http://www.pnas.org/content/108/17/6878.full.pdf} \BibitemShut {NoStop}%
	\bibitem [{\citenamefont {Yi}\ \emph {et~al.}(2012)\citenamefont {Yi},
		\citenamefont {Lu}, \citenamefont {Moore}, \citenamefont {Kihou},
		\citenamefont {Lee}, \citenamefont {Iyo}, \citenamefont {Eisaki},
		\citenamefont {Yoshida}, \citenamefont {Fujimori},\ and\ \citenamefont
		{Shen}}]{Yi2012}%
	\BibitemOpen
	\bibfield  {author} {\bibinfo {author} {\bibfnamefont {M}~\bibnamefont {Yi}},
		\bibinfo {author} {\bibfnamefont {D~H}\ \bibnamefont {Lu}}, \bibinfo {author}
		{\bibfnamefont {R~G}\ \bibnamefont {Moore}}, \bibinfo {author} {\bibfnamefont
			{K}~\bibnamefont {Kihou}}, \bibinfo {author} {\bibfnamefont {C-H}\
			\bibnamefont {Lee}}, \bibinfo {author} {\bibfnamefont {A}~\bibnamefont
			{Iyo}}, \bibinfo {author} {\bibfnamefont {H}~\bibnamefont {Eisaki}}, \bibinfo
		{author} {\bibfnamefont {T}~\bibnamefont {Yoshida}}, \bibinfo {author}
		{\bibfnamefont {A}~\bibnamefont {Fujimori}}, \ and\ \bibinfo {author}
		{\bibfnamefont {Z-X}\ \bibnamefont {Shen}},\ }\bibfield  {title} {\enquote
		{\bibinfo {title} {{Electronic reconstruction through the structural and
					magnetic transitions in detwinned NaFeAs}},}\ }\href
	{http://stacks.iop.org/1367-2630/14/i=7/a=073019} {\bibfield  {journal}
		{\bibinfo  {journal} {New Journal of Physics}\ }\textbf {\bibinfo {volume}
			{14}},\ \bibinfo {pages} {073019} (\bibinfo {year} {2012})}\BibitemShut
	{NoStop}%
	\bibitem [{\citenamefont {Zhang}\ \emph {et~al.}(2012)\citenamefont {Zhang},
		\citenamefont {He}, \citenamefont {Ye}, \citenamefont {Jiang}, \citenamefont
		{Chen}, \citenamefont {Xu}, \citenamefont {Ge}, \citenamefont {Xie},
		\citenamefont {Wei}, \citenamefont {Aeschlimann}, \citenamefont {Cui},
		\citenamefont {Shi}, \citenamefont {Hu},\ and\ \citenamefont
		{Feng}}]{Zhang2012b}%
	\BibitemOpen
	\bibfield  {author} {\bibinfo {author} {\bibfnamefont {Y.}~\bibnamefont
			{Zhang}}, \bibinfo {author} {\bibfnamefont {C.}~\bibnamefont {He}}, \bibinfo
		{author} {\bibfnamefont {Z.~R.}\ \bibnamefont {Ye}}, \bibinfo {author}
		{\bibfnamefont {J.}~\bibnamefont {Jiang}}, \bibinfo {author} {\bibfnamefont
			{F.}~\bibnamefont {Chen}}, \bibinfo {author} {\bibfnamefont {M.}~\bibnamefont
			{Xu}}, \bibinfo {author} {\bibfnamefont {Q.~Q.}\ \bibnamefont {Ge}}, \bibinfo
		{author} {\bibfnamefont {B.~P.}\ \bibnamefont {Xie}}, \bibinfo {author}
		{\bibfnamefont {J.}~\bibnamefont {Wei}}, \bibinfo {author} {\bibfnamefont
			{M.}~\bibnamefont {Aeschlimann}}, \bibinfo {author} {\bibfnamefont {X.~Y.}\
			\bibnamefont {Cui}}, \bibinfo {author} {\bibfnamefont {M.}~\bibnamefont
			{Shi}}, \bibinfo {author} {\bibfnamefont {J.~P.}\ \bibnamefont {Hu}}, \ and\
		\bibinfo {author} {\bibfnamefont {D.~L.}\ \bibnamefont {Feng}},\ }\bibfield
	{title} {\enquote {\bibinfo {title} {{Symmetry breaking via orbital-dependent
					reconstruction of electronic structure in detwinned NaFeAs}},}\ }\href
	{\doibase 10.1103/PhysRevB.85.085121} {\bibfield  {journal} {\bibinfo
			{journal} {Phys. Rev. B}\ }\textbf {\bibinfo {volume} {85}},\ \bibinfo
		{pages} {085121} (\bibinfo {year} {2012})}\BibitemShut {NoStop}%
	\bibitem [{\citenamefont {Wright}\ \emph {et~al.}(2012)\citenamefont {Wright},
		\citenamefont {Lancaster}, \citenamefont {Franke}, \citenamefont {Steele},
		\citenamefont {M{\"{o}}ller}, \citenamefont {Pitcher}, \citenamefont
		{Corkett}, \citenamefont {Parker}, \citenamefont {Free}, \citenamefont
		{Pratt}, \citenamefont {Baker}, \citenamefont {Clarke},\ and\ \citenamefont
		{Blundell}}]{Wright2012}%
	\BibitemOpen
	\bibfield  {author} {\bibinfo {author} {\bibfnamefont {J~D}\ \bibnamefont
			{Wright}}, \bibinfo {author} {\bibfnamefont {T}~\bibnamefont {Lancaster}},
		\bibinfo {author} {\bibfnamefont {I}~\bibnamefont {Franke}}, \bibinfo
		{author} {\bibfnamefont {A~J}\ \bibnamefont {Steele}}, \bibinfo {author}
		{\bibfnamefont {J~S}\ \bibnamefont {M{\"{o}}ller}}, \bibinfo {author}
		{\bibfnamefont {M~J}\ \bibnamefont {Pitcher}}, \bibinfo {author}
		{\bibfnamefont {A~J}\ \bibnamefont {Corkett}}, \bibinfo {author}
		{\bibfnamefont {D~R}\ \bibnamefont {Parker}}, \bibinfo {author}
		{\bibfnamefont {D~G}\ \bibnamefont {Free}}, \bibinfo {author} {\bibfnamefont
			{F~L}\ \bibnamefont {Pratt}}, \bibinfo {author} {\bibfnamefont {P~J}\
			\bibnamefont {Baker}}, \bibinfo {author} {\bibfnamefont {S~J}\ \bibnamefont
			{Clarke}}, \ and\ \bibinfo {author} {\bibfnamefont {S~J}\ \bibnamefont
			{Blundell}},\ }\bibfield  {title} {\enquote {\bibinfo {title} {{Gradual
					destruction of magnetism in the superconducting family
					NaFe$_{1-x}$Co$_x$As}},}\ }\href {\doibase 10.1103/PhysRevB.85.054503}
	{\bibfield  {journal} {\bibinfo  {journal} {Phys. Rev. B}\ }\textbf {\bibinfo
			{volume} {85}},\ \bibinfo {pages} {054503} (\bibinfo {year}
		{2012})}\BibitemShut {NoStop}%
	\bibitem [{\citenamefont {Fernandes}\ \emph {et~al.}(2014)\citenamefont
		{Fernandes}, \citenamefont {Chubukov},\ and\ \citenamefont
		{Schmalian}}]{Fernandes2014}%
	\BibitemOpen
	\bibfield  {author} {\bibinfo {author} {\bibfnamefont {R~M}\ \bibnamefont
			{Fernandes}}, \bibinfo {author} {\bibfnamefont {A~V}\ \bibnamefont
			{Chubukov}}, \ and\ \bibinfo {author} {\bibfnamefont {J}~\bibnamefont
			{Schmalian}},\ }\bibfield  {title} {\enquote {\bibinfo {title} {{What drives
					nematic order in iron-based superconductors?}}}\ }\href {\doibase
		10.1038/nphys2877} {\bibfield  {journal} {\bibinfo  {journal} {Nat. Phys.}\
		}\textbf {\bibinfo {volume} {10}},\ \bibinfo {pages} {97--104} (\bibinfo
		{year} {2014})}\BibitemShut {NoStop}%
	\bibitem [{\citenamefont {Rosenthal}\ \emph {et~al.}(2014)\citenamefont
		{Rosenthal}, \citenamefont {Andrade}, \citenamefont {Arguello}, \citenamefont
		{Fernandes}, \citenamefont {Xing}, \citenamefont {Wang}, \citenamefont {Jin},
		\citenamefont {Millis},\ and\ \citenamefont {Pasupathy}}]{Rosenthal2014}%
	\BibitemOpen
	\bibfield  {author} {\bibinfo {author} {\bibfnamefont {E.~P.}\ \bibnamefont
			{Rosenthal}}, \bibinfo {author} {\bibfnamefont {E.~F.}\ \bibnamefont
			{Andrade}}, \bibinfo {author} {\bibfnamefont {C.~J.}\ \bibnamefont
			{Arguello}}, \bibinfo {author} {\bibfnamefont {R.~M.}\ \bibnamefont
			{Fernandes}}, \bibinfo {author} {\bibfnamefont {L.~Y.}\ \bibnamefont {Xing}},
		\bibinfo {author} {\bibfnamefont {X.~C.}\ \bibnamefont {Wang}}, \bibinfo
		{author} {\bibfnamefont {C.~Q.}\ \bibnamefont {Jin}}, \bibinfo {author}
		{\bibfnamefont {A.~J.}\ \bibnamefont {Millis}}, \ and\ \bibinfo {author}
		{\bibfnamefont {A.~N.}\ \bibnamefont {Pasupathy}},\ }\bibfield  {title}
	{\enquote {\bibinfo {title} {{Visualization of electron nematicity and
					unidirectional antiferroic fluctuations at high temperatures in NaFeAs}},}\
	}\href {\doibase 10.1038/nphys2870} {\bibfield  {journal} {\bibinfo
			{journal} {Nat. Phys.}\ }\textbf {\bibinfo {volume} {10}},\ \bibinfo {pages}
		{225--232} (\bibinfo {year} {2014})}\BibitemShut {NoStop}%
	\bibitem [{\citenamefont {Deng}\ \emph {et~al.}(2015)\citenamefont {Deng},
		\citenamefont {Liu}, \citenamefont {Xing}, \citenamefont {Yang},\ and\
		\citenamefont {Wen}}]{Deng2015}%
	\BibitemOpen
	\bibfield  {author} {\bibinfo {author} {\bibfnamefont {Qiang}\ \bibnamefont
			{Deng}}, \bibinfo {author} {\bibfnamefont {Jianzhong}\ \bibnamefont {Liu}},
		\bibinfo {author} {\bibfnamefont {Jie}\ \bibnamefont {Xing}}, \bibinfo
		{author} {\bibfnamefont {Huan}\ \bibnamefont {Yang}}, \ and\ \bibinfo
		{author} {\bibfnamefont {Hai~Hu}\ \bibnamefont {Wen}},\ }\bibfield  {title}
	{\enquote {\bibinfo {title} {{Simultaneous vanishing of nematic electronic
					state and structural orthorhombicity in NaFe$_{1-x}$Co$_x$As single
					crystals}},}\ }\href {\doibase 10.1103/PhysRevB.91.020508} {\bibfield
		{journal} {\bibinfo  {journal} {Phys. Rev. B - Condens. Matter Mater. Phys.}\
		}\textbf {\bibinfo {volume} {91}},\ \bibinfo {pages} {020508(R)} (\bibinfo
		{year} {2015})}\BibitemShut {NoStop}%
	\bibitem [{\citenamefont {Parker}\ \emph {et~al.}(2010)\citenamefont {Parker},
		\citenamefont {Smith}, \citenamefont {Lancaster}, \citenamefont {Steele},
		\citenamefont {Franke}, \citenamefont {Baker}, \citenamefont {Pratt},
		\citenamefont {Pitcher}, \citenamefont {Blundell},\ and\ \citenamefont
		{Clarke}}]{Parker2010}%
	\BibitemOpen
	\bibfield  {author} {\bibinfo {author} {\bibfnamefont {Dinah~R.}\
			\bibnamefont {Parker}}, \bibinfo {author} {\bibfnamefont {Matthew J.~P.}\
			\bibnamefont {Smith}}, \bibinfo {author} {\bibfnamefont {Tom}\ \bibnamefont
			{Lancaster}}, \bibinfo {author} {\bibfnamefont {Andrew~J.}\ \bibnamefont
			{Steele}}, \bibinfo {author} {\bibfnamefont {Isabel}\ \bibnamefont {Franke}},
		\bibinfo {author} {\bibfnamefont {Peter~J.}\ \bibnamefont {Baker}}, \bibinfo
		{author} {\bibfnamefont {Francis~L.}\ \bibnamefont {Pratt}}, \bibinfo
		{author} {\bibfnamefont {Michael~J.}\ \bibnamefont {Pitcher}}, \bibinfo
		{author} {\bibfnamefont {Stephen~J.}\ \bibnamefont {Blundell}}, \ and\
		\bibinfo {author} {\bibfnamefont {Simon~J.}\ \bibnamefont {Clarke}},\
	}\bibfield  {title} {\enquote {\bibinfo {title} {{Control of the Competition
					between a Magnetic Phase and a Superconducting Phase in Cobalt-Doped and
					Nickel-Doped NaFeAs Using Electron Count}},}\ }\href {\doibase
		10.1103/PhysRevLett.104.057007} {\bibfield  {journal} {\bibinfo  {journal}
			{Phys. Rev. Lett.}\ }\textbf {\bibinfo {volume} {104}},\ \bibinfo {pages}
		{057007} (\bibinfo {year} {2010})}\BibitemShut {NoStop}%
	\bibitem [{\citenamefont {Steckel}\ \emph {et~al.}(2015)\citenamefont
		{Steckel}, \citenamefont {Roslova}, \citenamefont {Beck}, \citenamefont
		{Morozov}, \citenamefont {Aswartham}, \citenamefont {Evtushinsky},
		\citenamefont {Blum}, \citenamefont {Abdel-Hafiez}, \citenamefont {Bombor},
		\citenamefont {Maletz}, \citenamefont {Borisenko}, \citenamefont {Shevelkov},
		\citenamefont {Wolter}, \citenamefont {Hess}, \citenamefont {Wurmehl},\ and\
		\citenamefont {B\"uchner}}]{Steckel2015}%
	\BibitemOpen
	\bibfield  {author} {\bibinfo {author} {\bibfnamefont {Frank}\ \bibnamefont
			{Steckel}}, \bibinfo {author} {\bibfnamefont {Maria}\ \bibnamefont
			{Roslova}}, \bibinfo {author} {\bibfnamefont {Robert}\ \bibnamefont {Beck}},
		\bibinfo {author} {\bibfnamefont {Igor}\ \bibnamefont {Morozov}}, \bibinfo
		{author} {\bibfnamefont {Saicharan}\ \bibnamefont {Aswartham}}, \bibinfo
		{author} {\bibfnamefont {Daniil}\ \bibnamefont {Evtushinsky}}, \bibinfo
		{author} {\bibfnamefont {Christian G.~F.}\ \bibnamefont {Blum}}, \bibinfo
		{author} {\bibfnamefont {Mahmoud}\ \bibnamefont {Abdel-Hafiez}}, \bibinfo
		{author} {\bibfnamefont {Dirk}\ \bibnamefont {Bombor}}, \bibinfo {author}
		{\bibfnamefont {Janek}\ \bibnamefont {Maletz}}, \bibinfo {author}
		{\bibfnamefont {Sergey}\ \bibnamefont {Borisenko}}, \bibinfo {author}
		{\bibfnamefont {Andrei~V.}\ \bibnamefont {Shevelkov}}, \bibinfo {author}
		{\bibfnamefont {Anja U.~B.}\ \bibnamefont {Wolter}}, \bibinfo {author}
		{\bibfnamefont {Christian}\ \bibnamefont {Hess}}, \bibinfo {author}
		{\bibfnamefont {Sabine}\ \bibnamefont {Wurmehl}}, \ and\ \bibinfo {author}
		{\bibfnamefont {Bernd}\ \bibnamefont {B\"uchner}},\ }\bibfield  {title}
	{\enquote {\bibinfo {title} {{Crystal growth and electronic phase diagram of
					$4d\text{\ensuremath{-}}\mathrm{doped}$
					${\mathrm{Na}}_{1\ensuremath{-}\ensuremath{\delta}}{\mathrm{Fe}}_{1\ensuremath{-}x}{\mathrm{Rh}}_{x}\mathrm{As}$
					in comparison to
					$3d\ensuremath{-}\text{doped}~{\mathrm{Na}}_{1\ensuremath{-}\ensuremath{\delta}}{\mathrm{Fe}}_{1\ensuremath{-}x}{\mathrm{Co}}_{x}\mathrm{As}$}},}\
	}\href {\doibase 10.1103/PhysRevB.91.184516} {\bibfield  {journal} {\bibinfo
			{journal} {Phys. Rev. B}\ }\textbf {\bibinfo {volume} {91}},\ \bibinfo
		{pages} {184516} (\bibinfo {year} {2015})}\BibitemShut {NoStop}%
	\bibitem [{\citenamefont {Kim}\ \emph {et~al.}(2011)\citenamefont {Kim},
		\citenamefont {Fernandes}, \citenamefont {Kreyssig}, \citenamefont {Kim},
		\citenamefont {Thaler}, \citenamefont {Bud'ko}, \citenamefont {Canfield},
		\citenamefont {McQueeney}, \citenamefont {Schmalian},\ and\ \citenamefont
		{Goldman}}]{Kim2011}%
	\BibitemOpen
	\bibfield  {author} {\bibinfo {author} {\bibfnamefont {M.~G.}\ \bibnamefont
			{Kim}}, \bibinfo {author} {\bibfnamefont {R.~M.}\ \bibnamefont {Fernandes}},
		\bibinfo {author} {\bibfnamefont {A.}~\bibnamefont {Kreyssig}}, \bibinfo
		{author} {\bibfnamefont {J.~W.}\ \bibnamefont {Kim}}, \bibinfo {author}
		{\bibfnamefont {A.}~\bibnamefont {Thaler}}, \bibinfo {author} {\bibfnamefont
			{S.~L.}\ \bibnamefont {Bud'ko}}, \bibinfo {author} {\bibfnamefont {P.~C.}\
			\bibnamefont {Canfield}}, \bibinfo {author} {\bibfnamefont {R.~J.}\
			\bibnamefont {McQueeney}}, \bibinfo {author} {\bibfnamefont {J.}~\bibnamefont
			{Schmalian}}, \ and\ \bibinfo {author} {\bibfnamefont {A.~I.}\ \bibnamefont
			{Goldman}},\ }\bibfield  {title} {\enquote {\bibinfo {title} {{Character of
					the structural and magnetic phase transitions in the parent and
					electron-doped BaFe${}_{2}$As${}_{2}$ compounds}},}\ }\href {\doibase
		10.1103/PhysRevB.83.134522} {\bibfield  {journal} {\bibinfo  {journal} {Phys.
				Rev. B}\ }\textbf {\bibinfo {volume} {83}},\ \bibinfo {pages} {134522}
		(\bibinfo {year} {2011})}\BibitemShut {NoStop}%
	\bibitem [{\citenamefont {Zhang}\ \emph {et~al.}(2015)\citenamefont {Zhang},
		\citenamefont {Fernandes}, \citenamefont {Lamsal}, \citenamefont {Yan},
		\citenamefont {Chi}, \citenamefont {Tucker}, \citenamefont {Pratt},
		\citenamefont {Lynn}, \citenamefont {McCallum}, \citenamefont {Canfield},
		\citenamefont {Lograsso}, \citenamefont {Goldman}, \citenamefont {Vaknin},\
		and\ \citenamefont {McQueeney}}]{Zhang2015PRL}%
	\BibitemOpen
	\bibfield  {author} {\bibinfo {author} {\bibfnamefont {Qiang}\ \bibnamefont
			{Zhang}}, \bibinfo {author} {\bibfnamefont {Rafael~M.}\ \bibnamefont
			{Fernandes}}, \bibinfo {author} {\bibfnamefont {Jagat}\ \bibnamefont
			{Lamsal}}, \bibinfo {author} {\bibfnamefont {Jiaqiang}\ \bibnamefont {Yan}},
		\bibinfo {author} {\bibfnamefont {Songxue}\ \bibnamefont {Chi}}, \bibinfo
		{author} {\bibfnamefont {Gregory~S.}\ \bibnamefont {Tucker}}, \bibinfo
		{author} {\bibfnamefont {Daniel~K.}\ \bibnamefont {Pratt}}, \bibinfo {author}
		{\bibfnamefont {Jeffrey~W.}\ \bibnamefont {Lynn}}, \bibinfo {author}
		{\bibfnamefont {R.~W.}\ \bibnamefont {McCallum}}, \bibinfo {author}
		{\bibfnamefont {Paul~C.}\ \bibnamefont {Canfield}}, \bibinfo {author}
		{\bibfnamefont {Thomas~A.}\ \bibnamefont {Lograsso}}, \bibinfo {author}
		{\bibfnamefont {Alan~I.}\ \bibnamefont {Goldman}}, \bibinfo {author}
		{\bibfnamefont {David}\ \bibnamefont {Vaknin}}, \ and\ \bibinfo {author}
		{\bibfnamefont {Robert~J.}\ \bibnamefont {McQueeney}},\ }\bibfield  {title}
	{\enquote {\bibinfo {title} {{Neutron-Scattering Measurements of Spin
					Excitations in LaFeAsO and
					$\mathrm{Ba}({\mathrm{Fe}}_{0.953}{\mathrm{Co}}_{0.047}{)}_{2}{\mathrm{As}}_{2}$:
					Evidence for a Sharp Enhancement of Spin Fluctuations by Nematic Order}},}\
	}\href {\doibase 10.1103/PhysRevLett.114.057001} {\bibfield  {journal}
		{\bibinfo  {journal} {Phys. Rev. Lett.}\ }\textbf {\bibinfo {volume} {114}},\
		\bibinfo {pages} {057001} (\bibinfo {year} {2015})}\BibitemShut {NoStop}%
	\bibitem [{\citenamefont {Borisenko}\ \emph {et~al.}(2010)\citenamefont
		{Borisenko}, \citenamefont {Zabolotnyy}, \citenamefont {Evtushinsky},
		\citenamefont {Kim}, \citenamefont {Morozov}, \citenamefont {Yaresko},
		\citenamefont {Kordyuk}, \citenamefont {Behr}, \citenamefont {Vasiliev},
		\citenamefont {Follath},\ and\ \citenamefont {B\"uchner}}]{Borisenko2010}%
	\BibitemOpen
	\bibfield  {author} {\bibinfo {author} {\bibfnamefont {S.~V.}\ \bibnamefont
			{Borisenko}}, \bibinfo {author} {\bibfnamefont {V.~B.}\ \bibnamefont
			{Zabolotnyy}}, \bibinfo {author} {\bibfnamefont {D.~V.}\ \bibnamefont
			{Evtushinsky}}, \bibinfo {author} {\bibfnamefont {T.~K.}\ \bibnamefont
			{Kim}}, \bibinfo {author} {\bibfnamefont {I.~V.}\ \bibnamefont {Morozov}},
		\bibinfo {author} {\bibfnamefont {A.~N.}\ \bibnamefont {Yaresko}}, \bibinfo
		{author} {\bibfnamefont {A.~A.}\ \bibnamefont {Kordyuk}}, \bibinfo {author}
		{\bibfnamefont {G.}~\bibnamefont {Behr}}, \bibinfo {author} {\bibfnamefont
			{A.}~\bibnamefont {Vasiliev}}, \bibinfo {author} {\bibfnamefont
			{R.}~\bibnamefont {Follath}}, \ and\ \bibinfo {author} {\bibfnamefont
			{B.}~\bibnamefont {B\"uchner}},\ }\bibfield  {title} {\enquote {\bibinfo
			{title} {{Superconductivity without Nesting in LiFeAs}},}\ }\href {\doibase
		10.1103/PhysRevLett.105.067002} {\bibfield  {journal} {\bibinfo  {journal}
			{Phys. Rev. Lett.}\ }\textbf {\bibinfo {volume} {105}},\ \bibinfo {pages}
		{067002} (\bibinfo {year} {2010})}\BibitemShut {NoStop}%
	\bibitem [{\citenamefont {Borisenko}\ \emph {et~al.}(2016)\citenamefont
		{Borisenko}, \citenamefont {Evtushinsky}, \citenamefont {Liu}, \citenamefont
		{Morozov}, \citenamefont {Kappenberger}, \citenamefont {Wurmehl},
		\citenamefont {Buchner}, \citenamefont {Yaresko}, \citenamefont {Kim},
		\citenamefont {Hoesch}, \citenamefont {Wolf},\ and\ \citenamefont
		{Zhigadlo}}]{Borisenko2015}%
	\BibitemOpen
	\bibfield  {author} {\bibinfo {author} {\bibfnamefont {S~V}\ \bibnamefont
			{Borisenko}}, \bibinfo {author} {\bibfnamefont {D~V}\ \bibnamefont
			{Evtushinsky}}, \bibinfo {author} {\bibfnamefont {Z.-H.}\ \bibnamefont
			{Liu}}, \bibinfo {author} {\bibfnamefont {I}~\bibnamefont {Morozov}},
		\bibinfo {author} {\bibfnamefont {R}~\bibnamefont {Kappenberger}}, \bibinfo
		{author} {\bibfnamefont {S}~\bibnamefont {Wurmehl}}, \bibinfo {author}
		{\bibfnamefont {B}~\bibnamefont {Buchner}}, \bibinfo {author} {\bibfnamefont
			{A~N}\ \bibnamefont {Yaresko}}, \bibinfo {author} {\bibfnamefont {T~K}\
			\bibnamefont {Kim}}, \bibinfo {author} {\bibfnamefont {M}~\bibnamefont
			{Hoesch}}, \bibinfo {author} {\bibfnamefont {T}~\bibnamefont {Wolf}}, \ and\
		\bibinfo {author} {\bibfnamefont {N~D}\ \bibnamefont {Zhigadlo}},\ }\bibfield
	{title} {\enquote {\bibinfo {title} {{Direct observation of spin-orbit
					coupling in iron-based superconductors}},}\ }\href
	{http://dx.doi.org/10.1038/nphys3594 10.1038/nphys3594
		http://www.nature.com/nphys/journal/vaop/ncurrent/abs/nphys3594.html{\#}supplementary-information}
	{\bibfield  {journal} {\bibinfo  {journal} {Nat Phys}\ }\textbf {\bibinfo
			{volume} {12}},\ \bibinfo {pages} {311--317} (\bibinfo {year}
		{2016})}\BibitemShut {NoStop}%
	\bibitem [{\citenamefont {Yi}\ \emph {et~al.}(2017)\citenamefont {Yi},
		\citenamefont {Zhang}, \citenamefont {Shen},\ and\ \citenamefont
		{Lu}}]{Yi2017_arxiv}%
	\BibitemOpen
	\bibfield  {author} {\bibinfo {author} {\bibfnamefont {M}~\bibnamefont {Yi}},
		\bibinfo {author} {\bibfnamefont {Y}~\bibnamefont {Zhang}}, \bibinfo {author}
		{\bibfnamefont {Z.-X.}\ \bibnamefont {Shen}}, \ and\ \bibinfo {author}
		{\bibfnamefont {D}~\bibnamefont {Lu}},\ }\bibfield  {title} {\enquote
		{\bibinfo {title} {{Role of the orbital degree of freedom in iron-based
					superconductors}},}\ }\href {https://arxiv.org/abs/1703.08622} {\  (\bibinfo
		{year} {2017})},\ \Eprint {http://arxiv.org/abs/1703.08622}
	{arXiv:1703.08622} \BibitemShut {NoStop}%
	\bibitem [{\citenamefont {Watson}\ \emph
		{et~al.}(2017{\natexlab{a}})\citenamefont {Watson}, \citenamefont
		{Haghighirad}, \citenamefont {Rhodes}, \citenamefont {Hoesch},\ and\
		\citenamefont {Kim}}]{Watson2017d_arxiv}%
	\BibitemOpen
	\bibfield  {author} {\bibinfo {author} {\bibfnamefont {M.D.}\ \bibnamefont
			{Watson}}, \bibinfo {author} {\bibfnamefont {A.~A.}\ \bibnamefont
			{Haghighirad}}, \bibinfo {author} {\bibfnamefont {L.~C.}\ \bibnamefont
			{Rhodes}}, \bibinfo {author} {\bibfnamefont {M}~\bibnamefont {Hoesch}}, \
		and\ \bibinfo {author} {\bibfnamefont {T.~K.}\ \bibnamefont {Kim}},\
	}\bibfield  {title} {\enquote {\bibinfo {title} {{Electronic anisotropies
					revealed by detwinned ARPES measurements of FeSe}},}\ }\href
	{http://iopscience.iop.org/article/10.1088/1367-2630/aa8a04} {\bibfield
		{journal} {\bibinfo  {journal} {New Journal of Physics (in press)}\ }
		(\bibinfo {year} {2017}{\natexlab{a}})}\BibitemShut {NoStop}%
	\bibitem [{\citenamefont {Hoesch}\ \emph {et~al.}(2017)\citenamefont {Hoesch},
		\citenamefont {Kim}, \citenamefont {Dudin}, \citenamefont {Wang},
		\citenamefont {Scott}, \citenamefont {Harris}, \citenamefont {Patel},
		\citenamefont {Matthews}, \citenamefont {Hawkins}, \citenamefont {Alcock},
		\citenamefont {Richter}, \citenamefont {Mudd}, \citenamefont {Basham},
		\citenamefont {Pratt}, \citenamefont {Leicester}, \citenamefont {Longhi},
		\citenamefont {Tamai},\ and\ \citenamefont {Baumberger}}]{I05beamlinepaper}%
	\BibitemOpen
	\bibfield  {author} {\bibinfo {author} {\bibfnamefont {M.}~\bibnamefont
			{Hoesch}}, \bibinfo {author} {\bibfnamefont {T.~K.}\ \bibnamefont {Kim}},
		\bibinfo {author} {\bibfnamefont {P.}~\bibnamefont {Dudin}}, \bibinfo
		{author} {\bibfnamefont {H.}~\bibnamefont {Wang}}, \bibinfo {author}
		{\bibfnamefont {S.}~\bibnamefont {Scott}}, \bibinfo {author} {\bibfnamefont
			{P.}~\bibnamefont {Harris}}, \bibinfo {author} {\bibfnamefont
			{S.}~\bibnamefont {Patel}}, \bibinfo {author} {\bibfnamefont
			{M.}~\bibnamefont {Matthews}}, \bibinfo {author} {\bibfnamefont
			{D.}~\bibnamefont {Hawkins}}, \bibinfo {author} {\bibfnamefont {S.~G.}\
			\bibnamefont {Alcock}}, \bibinfo {author} {\bibfnamefont {T.}~\bibnamefont
			{Richter}}, \bibinfo {author} {\bibfnamefont {J.~J.}\ \bibnamefont {Mudd}},
		\bibinfo {author} {\bibfnamefont {M.}~\bibnamefont {Basham}}, \bibinfo
		{author} {\bibfnamefont {L.}~\bibnamefont {Pratt}}, \bibinfo {author}
		{\bibfnamefont {P.}~\bibnamefont {Leicester}}, \bibinfo {author}
		{\bibfnamefont {E.~C.}\ \bibnamefont {Longhi}}, \bibinfo {author}
		{\bibfnamefont {A.}~\bibnamefont {Tamai}}, \ and\ \bibinfo {author}
		{\bibfnamefont {F.}~\bibnamefont {Baumberger}},\ }\bibfield  {title}
	{\enquote {\bibinfo {title} {A facility for the analysis of the electronic
				structures of solids and their surfaces by synchrotron radiation
				photoelectron spectroscopy},}\ }\href {http://dx.doi.org/10.1063/1.4973562}
	{\bibfield  {journal} {\bibinfo  {journal} {Review of Scientific
				Instruments}\ }\textbf {\bibinfo {volume} {88}},\ \bibinfo {pages} {013106}
		(\bibinfo {year} {2017})}\BibitemShut {NoStop}%
	\bibitem [{Note1()}]{Note1}%
	\BibitemOpen
	\bibinfo {note} {Formally there are four possible domains in the magnetic
		phase but for the purposes of ARPES measurements we can simply consider only
		two possible domain orientations.}\BibitemShut {Stop}%
	\bibitem [{\citenamefont {Fisher}\ \emph {et~al.}(2011)\citenamefont {Fisher},
		\citenamefont {Degiorgi},\ and\ \citenamefont {Shen}}]{Fisher2011a}%
	\BibitemOpen
	\bibfield  {author} {\bibinfo {author} {\bibfnamefont {I~R}\ \bibnamefont
			{Fisher}}, \bibinfo {author} {\bibfnamefont {L}~\bibnamefont {Degiorgi}}, \
		and\ \bibinfo {author} {\bibfnamefont {Z~X}\ \bibnamefont {Shen}},\
	}\bibfield  {title} {\enquote {\bibinfo {title} {{In-plane electronic
					anisotropy of underdoped `122' Fe-arsenide superconductors revealed by
					measurements of detwinned single crystals}},}\ }\href
	{http://stacks.iop.org/0034-4885/74/i=12/a=124506} {\bibfield  {journal}
		{\bibinfo  {journal} {Reports on Progress in Physics}\ }\textbf {\bibinfo
			{volume} {74}},\ \bibinfo {pages} {124506} (\bibinfo {year}
		{2011})}\BibitemShut {NoStop}%
	\bibitem [{Note2()}]{Note2}%
	\BibitemOpen
	\bibinfo {note} {The detwinned data have slightly compromised resolution
		compared with the unstrained sample; this is partially due to necessarily
		shorter measurement times on the detwinned samples and possibly extrinsic
		issues due to the more complex sample mounting environment}\BibitemShut
	{NoStop}%
	\bibitem [{\citenamefont {Chen}\ \emph {et~al.}(2015)\citenamefont {Chen},
		\citenamefont {Maiti}, \citenamefont {Linscheid},\ and\ \citenamefont
		{Hirschfeld}}]{Chen2015}%
	\BibitemOpen
	\bibfield  {author} {\bibinfo {author} {\bibfnamefont {Xiao}\ \bibnamefont
			{Chen}}, \bibinfo {author} {\bibfnamefont {S}~\bibnamefont {Maiti}}, \bibinfo
		{author} {\bibfnamefont {A}~\bibnamefont {Linscheid}}, \ and\ \bibinfo
		{author} {\bibfnamefont {P~J}\ \bibnamefont {Hirschfeld}},\ }\bibfield
	{title} {\enquote {\bibinfo {title} {{Electron pairing in the presence of
					incipient bands in iron-based superconductors}},}\ }\href {\doibase
		10.1103/PhysRevB.92.224514} {\bibfield  {journal} {\bibinfo  {journal} {Phys.
				Rev. B}\ }\textbf {\bibinfo {volume} {92}},\ \bibinfo {pages} {224514}
		(\bibinfo {year} {2015})}\BibitemShut {NoStop}%
	\bibitem [{\citenamefont {Jiang}\ \emph {et~al.}(2016)\citenamefont {Jiang},
		\citenamefont {Hu}, \citenamefont {Ding},\ and\ \citenamefont
		{Wang}}]{Jiang2016}%
	\BibitemOpen
	\bibfield  {author} {\bibinfo {author} {\bibfnamefont {Kun}\ \bibnamefont
			{Jiang}}, \bibinfo {author} {\bibfnamefont {Jiangping}\ \bibnamefont {Hu}},
		\bibinfo {author} {\bibfnamefont {Hong}\ \bibnamefont {Ding}}, \ and\
		\bibinfo {author} {\bibfnamefont {Ziqiang}\ \bibnamefont {Wang}},\ }\bibfield
	{title} {\enquote {\bibinfo {title} {{Interatomic Coulomb interaction and
					electron nematic bond order in FeSe}},}\ }\href {\doibase
		10.1103/PhysRevB.93.115138} {\bibfield  {journal} {\bibinfo  {journal} {Phys.
				Rev. B}\ }\textbf {\bibinfo {volume} {93}},\ \bibinfo {pages} {115138}
		(\bibinfo {year} {2016})}\BibitemShut {NoStop}%
	\bibitem [{\citenamefont {Ortenzi}\ \emph {et~al.}(2009)\citenamefont
		{Ortenzi}, \citenamefont {Cappelluti}, \citenamefont {Benfatto},\ and\
		\citenamefont {Pietronero}}]{Ortenzi2009}%
	\BibitemOpen
	\bibfield  {author} {\bibinfo {author} {\bibfnamefont {L.}~\bibnamefont
			{Ortenzi}}, \bibinfo {author} {\bibfnamefont {E.}~\bibnamefont {Cappelluti}},
		\bibinfo {author} {\bibfnamefont {L.}~\bibnamefont {Benfatto}}, \ and\
		\bibinfo {author} {\bibfnamefont {L.}~\bibnamefont {Pietronero}},\ }\bibfield
	{title} {\enquote {\bibinfo {title} {{Fermi-Surface Shrinking and Interband
					Coupling in Iron-Based Pnictides}},}\ }\href {\doibase
		10.1103/PhysRevLett.103.046404} {\bibfield  {journal} {\bibinfo  {journal}
			{Phys. Rev. Lett.}\ }\textbf {\bibinfo {volume} {103}},\ \bibinfo {pages}
		{046404} (\bibinfo {year} {2009})}\BibitemShut {NoStop}%
	\bibitem [{\citenamefont {Charnukha}\ \emph {et~al.}(2016)\citenamefont
		{Charnukha}, \citenamefont {Post}, \citenamefont {Thirupathaiah},
		\citenamefont {Pr{\"{o}}pper}, \citenamefont {Wurmehl}, \citenamefont
		{Roslova}, \citenamefont {Morozov}, \citenamefont {B{\"{u}}chner},
		\citenamefont {Yaresko}, \citenamefont {Boris}, \citenamefont {Borisenko},\
		and\ \citenamefont {Basov}}]{Charnukha2016}%
	\BibitemOpen
	\bibfield  {author} {\bibinfo {author} {\bibfnamefont {A}~\bibnamefont
			{Charnukha}}, \bibinfo {author} {\bibfnamefont {K~W}\ \bibnamefont {Post}},
		\bibinfo {author} {\bibfnamefont {S}~\bibnamefont {Thirupathaiah}}, \bibinfo
		{author} {\bibfnamefont {D}~\bibnamefont {Pr{\"{o}}pper}}, \bibinfo {author}
		{\bibfnamefont {S}~\bibnamefont {Wurmehl}}, \bibinfo {author} {\bibfnamefont
			{M}~\bibnamefont {Roslova}}, \bibinfo {author} {\bibfnamefont
			{I}~\bibnamefont {Morozov}}, \bibinfo {author} {\bibfnamefont
			{B}~\bibnamefont {B{\"{u}}chner}}, \bibinfo {author} {\bibfnamefont {A~N}\
			\bibnamefont {Yaresko}}, \bibinfo {author} {\bibfnamefont {A~V}\ \bibnamefont
			{Boris}}, \bibinfo {author} {\bibfnamefont {S~V}\ \bibnamefont {Borisenko}},
		\ and\ \bibinfo {author} {\bibfnamefont {D~N}\ \bibnamefont {Basov}},\
	}\bibfield  {title} {\enquote {\bibinfo {title} {{Weak-coupling
					superconductivity in a strongly correlated iron pnictide}},}\ }\href
	{\doibase 10.1038/srep18620} {\bibfield  {journal} {\bibinfo  {journal} {Sci.
				Rep.}\ }\textbf {\bibinfo {volume} {6}},\ \bibinfo {pages} {18620} (\bibinfo
		{year} {2016})}\BibitemShut {NoStop}%
	\bibitem [{\citenamefont {Evtushinsky}\ \emph {et~al.}(2017)\citenamefont
		{Evtushinsky}, \citenamefont {Yaresko}, \citenamefont {Zabolotnyy},
		\citenamefont {Maletz}, \citenamefont {Kim}, \citenamefont {Kordyuk},
		\citenamefont {Viazovska}, \citenamefont {Roslova}, \citenamefont {Morozov},
		\citenamefont {Beck}, \citenamefont {Aswartham}, \citenamefont {Harnagea},
		\citenamefont {Wurmehl}, \citenamefont {Berger}, \citenamefont {Rogalev},
		\citenamefont {Strocov}, \citenamefont {Wolf}, \citenamefont {Zhigadlo},
		\citenamefont {B\"uchner},\ and\ \citenamefont
		{Borisenko}}]{Evtushinsky2017NaFeAs}%
	\BibitemOpen
	\bibfield  {author} {\bibinfo {author} {\bibfnamefont {D.~V.}\ \bibnamefont
			{Evtushinsky}}, \bibinfo {author} {\bibfnamefont {A.~N.}\ \bibnamefont
			{Yaresko}}, \bibinfo {author} {\bibfnamefont {V.~B.}\ \bibnamefont
			{Zabolotnyy}}, \bibinfo {author} {\bibfnamefont {J.}~\bibnamefont {Maletz}},
		\bibinfo {author} {\bibfnamefont {T.~K.}\ \bibnamefont {Kim}}, \bibinfo
		{author} {\bibfnamefont {A.~A.}\ \bibnamefont {Kordyuk}}, \bibinfo {author}
		{\bibfnamefont {M.~S.}\ \bibnamefont {Viazovska}}, \bibinfo {author}
		{\bibfnamefont {M.}~\bibnamefont {Roslova}}, \bibinfo {author} {\bibfnamefont
			{I.}~\bibnamefont {Morozov}}, \bibinfo {author} {\bibfnamefont
			{R.}~\bibnamefont {Beck}}, \bibinfo {author} {\bibfnamefont {S.}~\bibnamefont
			{Aswartham}}, \bibinfo {author} {\bibfnamefont {L.}~\bibnamefont {Harnagea}},
		\bibinfo {author} {\bibfnamefont {S.}~\bibnamefont {Wurmehl}}, \bibinfo
		{author} {\bibfnamefont {H.}~\bibnamefont {Berger}}, \bibinfo {author}
		{\bibfnamefont {V.~A.}\ \bibnamefont {Rogalev}}, \bibinfo {author}
		{\bibfnamefont {V.~N.}\ \bibnamefont {Strocov}}, \bibinfo {author}
		{\bibfnamefont {T.}~\bibnamefont {Wolf}}, \bibinfo {author} {\bibfnamefont
			{N.~D.}\ \bibnamefont {Zhigadlo}}, \bibinfo {author} {\bibfnamefont
			{B.}~\bibnamefont {B\"uchner}}, \ and\ \bibinfo {author} {\bibfnamefont
			{S.~V.}\ \bibnamefont {Borisenko}},\ }\bibfield  {title} {\enquote {\bibinfo
			{title} {High-energy electronic interaction in the $3d$ band of
				high-temperature iron-based superconductors},}\ }\href {\doibase
		10.1103/PhysRevB.96.060501} {\bibfield  {journal} {\bibinfo  {journal} {Phys.
				Rev. B}\ }\textbf {\bibinfo {volume} {96}},\ \bibinfo {pages} {060501}
		(\bibinfo {year} {2017})}\BibitemShut {NoStop}%
	\bibitem [{\citenamefont {Nekrasov}\ \emph {et~al.}(2015)\citenamefont
		{Nekrasov}, \citenamefont {Pavlov},\ and\ \citenamefont
		{Sadovskii}}]{Nekrasov2015}%
	\BibitemOpen
	\bibfield  {author} {\bibinfo {author} {\bibfnamefont {I~A}\ \bibnamefont
			{Nekrasov}}, \bibinfo {author} {\bibfnamefont {N~S}\ \bibnamefont {Pavlov}},
		\ and\ \bibinfo {author} {\bibfnamefont {M~V}\ \bibnamefont {Sadovskii}},\
	}\bibfield  {title} {\enquote {\bibinfo {title} {{Electronic structure of
					NaFeAs superconductor: LDA+DMFT calculations compared to the ARPES
					experiment}},}\ }\href {\doibase 10.1134/S0021364015130123} {\bibfield
		{journal} {\bibinfo  {journal} {JETP Lett.}\ }\textbf {\bibinfo {volume}
			{102}},\ \bibinfo {pages} {26--31} (\bibinfo {year} {2015})}\BibitemShut
	{NoStop}%
	\bibitem [{\citenamefont {{Evtushinsky}}\ \emph {et~al.}(2016)\citenamefont
		{{Evtushinsky}}, \citenamefont {{Aichhorn}}, \citenamefont {{Sassa}},
		\citenamefont {{Liu}}, \citenamefont {{Maletz}}, \citenamefont {{Wolf}},
		\citenamefont {{Yaresko}}, \citenamefont {{Biermann}}, \citenamefont
		{{Borisenko}},\ and\ \citenamefont {{Buchner}}}]{Evtushinsky2017_arxiv}%
	\BibitemOpen
	\bibfield  {author} {\bibinfo {author} {\bibfnamefont {D.~V.}\ \bibnamefont
			{{Evtushinsky}}}, \bibinfo {author} {\bibfnamefont {M.}~\bibnamefont
			{{Aichhorn}}}, \bibinfo {author} {\bibfnamefont {Y.}~\bibnamefont {{Sassa}}},
		\bibinfo {author} {\bibfnamefont {Z.-H.}\ \bibnamefont {{Liu}}}, \bibinfo
		{author} {\bibfnamefont {J.}~\bibnamefont {{Maletz}}}, \bibinfo {author}
		{\bibfnamefont {T.}~\bibnamefont {{Wolf}}}, \bibinfo {author} {\bibfnamefont
			{A.~N.}\ \bibnamefont {{Yaresko}}}, \bibinfo {author} {\bibfnamefont
			{S.}~\bibnamefont {{Biermann}}}, \bibinfo {author} {\bibfnamefont {S.~V.}\
			\bibnamefont {{Borisenko}}}, \ and\ \bibinfo {author} {\bibfnamefont
			{B.}~\bibnamefont {{Buchner}}},\ }\bibfield  {title} {\enquote {\bibinfo
			{title} {{Direct observation of dispersive lower Hubbard band in iron-based
					superconductor FeSe}},}\ }\href {https://arxiv.org/abs/1612.02313} {\bibfield
		{journal} {\bibinfo  {journal} {arXiv:1612.02313}\ } (\bibinfo {year}
		{2016})}\BibitemShut {NoStop}%
	\bibitem [{\citenamefont {Watson}\ \emph
		{et~al.}(2017{\natexlab{b}})\citenamefont {Watson}, \citenamefont {Backes},
		\citenamefont {Haghighirad}, \citenamefont {Hoesch}, \citenamefont {Kim},
		\citenamefont {Coldea},\ and\ \citenamefont {Valent\'{i}}}]{Watson2017c}%
	\BibitemOpen
	\bibfield  {author} {\bibinfo {author} {\bibfnamefont {Matthew~D}\
			\bibnamefont {Watson}}, \bibinfo {author} {\bibfnamefont {Steffen}\
			\bibnamefont {Backes}}, \bibinfo {author} {\bibfnamefont {Amir~A}\
			\bibnamefont {Haghighirad}}, \bibinfo {author} {\bibfnamefont {Moritz}\
			\bibnamefont {Hoesch}}, \bibinfo {author} {\bibfnamefont {Timur~K}\
			\bibnamefont {Kim}}, \bibinfo {author} {\bibfnamefont {Amalia~I}\
			\bibnamefont {Coldea}}, \ and\ \bibinfo {author} {\bibfnamefont {Roser}\
			\bibnamefont {Valent\'{i}}},\ }\bibfield  {title} {\enquote {\bibinfo {title}
			{{Formation of Hubbard-like bands as a fingerprint of strong
					electron-electron interactions in FeSe}},}\ }\href {\doibase
		10.1103/PhysRevB.95.081106} {\bibfield  {journal} {\bibinfo  {journal} {Phys.
				Rev. B}\ }\textbf {\bibinfo {volume} {95}},\ \bibinfo {pages} {081106}
		(\bibinfo {year} {2017}{\natexlab{b}})}\BibitemShut {NoStop}%
	\bibitem [{\citenamefont {Brouet}\ \emph {et~al.}(2012)\citenamefont {Brouet},
		\citenamefont {Jensen}, \citenamefont {Lin}, \citenamefont {Taleb-Ibrahimi},
		\citenamefont {{Le F{\`{e}}vre}}, \citenamefont {Bertran}, \citenamefont
		{Lin}, \citenamefont {Ku}, \citenamefont {Forget},\ and\ \citenamefont
		{Colson}}]{Brouet2012}%
	\BibitemOpen
	\bibfield  {author} {\bibinfo {author} {\bibfnamefont {V.}~\bibnamefont
			{Brouet}}, \bibinfo {author} {\bibfnamefont {M.~F.}\ \bibnamefont
			{Jensen}}, \bibinfo {author} {\bibfnamefont {Ping~Hui}\ \bibnamefont {Lin}},
		\bibinfo {author} {\bibfnamefont {A.}~\bibnamefont {Taleb-Ibrahimi}},
		\bibinfo {author} {\bibfnamefont {P.}~\bibnamefont {{Le F{\`{e}}vre}}},
		\bibinfo {author} {\bibfnamefont {F.}~\bibnamefont {Bertran}}, \bibinfo
		{author} {\bibfnamefont {Chia~Hui}\ \bibnamefont {Lin}}, \bibinfo {author}
		{\bibfnamefont {Wei}\ \bibnamefont {Ku}}, \bibinfo {author} {\bibfnamefont
			{A.}~\bibnamefont {Forget}}, \ and\ \bibinfo {author} {\bibfnamefont
			{D.}~\bibnamefont {Colson}},\ }\bibfield  {title} {\enquote {\bibinfo {title}
			{{Impact of the two Fe unit cell on the electronic structure measured by
					ARPES in iron pnictides}},}\ }\href {\doibase 10.1103/PhysRevB.86.075123}
	{\bibfield  {journal} {\bibinfo  {journal} {Phys. Rev. B}\ }\textbf {\bibinfo
			{volume} {86}},\ \bibinfo {pages} {075123} (\bibinfo {year}
		{2012})}\BibitemShut {NoStop}%
	\bibitem [{\citenamefont {Watson}\ \emph {et~al.}(2016)\citenamefont {Watson},
		\citenamefont {Kim}, \citenamefont {Rhodes}, \citenamefont {Eschrig},
		\citenamefont {Hoesch}, \citenamefont {Haghighirad},\ and\ \citenamefont
		{Coldea}}]{Watson2016}%
	\BibitemOpen
	\bibfield  {author} {\bibinfo {author} {\bibfnamefont {M.~D.}\ \bibnamefont
			{Watson}}, \bibinfo {author} {\bibfnamefont {T.~K.}\ \bibnamefont {Kim}},
		\bibinfo {author} {\bibfnamefont {L.~C.}\ \bibnamefont {Rhodes}}, \bibinfo
		{author} {\bibfnamefont {M.}~\bibnamefont {Eschrig}}, \bibinfo {author}
		{\bibfnamefont {M.}~\bibnamefont {Hoesch}}, \bibinfo {author} {\bibfnamefont
			{A.~A.}\ \bibnamefont {Haghighirad}}, \ and\ \bibinfo {author} {\bibfnamefont
			{A.~I.}\ \bibnamefont {Coldea}},\ }\bibfield  {title} {\enquote {\bibinfo
			{title} {{Evidence for unidirectional nematic bond ordering in FeSe}},}\
	}\href {\doibase 10.1103/PhysRevB.94.201107} {\bibfield  {journal} {\bibinfo
			{journal} {Phys. Rev. B}\ }\textbf {\bibinfo {volume} {94}},\ \bibinfo
		{pages} {201107} (\bibinfo {year} {2016})}\BibitemShut {NoStop}%
	\bibitem [{\citenamefont {Chu}\ \emph {et~al.}(2012)\citenamefont {Chu},
		\citenamefont {Kuo}, \citenamefont {Analytis},\ and\ \citenamefont
		{Fisher}}]{Chu2012}%
	\BibitemOpen
	\bibfield  {author} {\bibinfo {author} {\bibfnamefont {Jiun-Haw}\
			\bibnamefont {Chu}}, \bibinfo {author} {\bibfnamefont {Hsueh-Hui}\
			\bibnamefont {Kuo}}, \bibinfo {author} {\bibfnamefont {James~G.}\
			\bibnamefont {Analytis}}, \ and\ \bibinfo {author} {\bibfnamefont {Ian~R.}\
			\bibnamefont {Fisher}},\ }\bibfield  {title} {\enquote {\bibinfo {title}
			{Divergent nematic susceptibility in an iron arsenide superconductor},}\
	}\href {\doibase 10.1126/science.1221713} {\bibfield  {journal} {\bibinfo
			{journal} {Science}\ }\textbf {\bibinfo {volume} {337}},\ \bibinfo {pages}
		{710--712} (\bibinfo {year} {2012})}\BibitemShut {NoStop}%
	\bibitem [{Note3()}]{Note3}%
	\BibitemOpen
	\bibinfo {note} {All data in Fig.~1 are taken with 42~eV LH photons in
		equivalent measurement geometries, only the detwinned sample is rotated by
		90$^\circ $. Previous studies of NaFeAs did not rotate the detwinned sample,
		but rather changed the polarisation of the light and the orientation of the
		measurement to achieve similar, but not exactly equivalent measurements along
		$\Gamma -\protect \mathrm {\protect \mathaccentV {bar}016{M}_Y}$ and $\Gamma
		-\protect \mathrm {\protect \mathaccentV {bar}016{M}_X}$.}\BibitemShut
	{Stop}%
	\bibitem [{Note4()}]{Note4}%
	\BibitemOpen
	\bibinfo {note} {See Supplemental Material at [URL will be inserted by
		publisher] for [Further ARPES spectra and discussion].}\BibitemShut {Stop}%
	\bibitem [{Note5()}]{Note5}%
	\BibitemOpen
	\bibinfo {note} {In He \protect \textit {et al.} \cite {He2010} it was
		claimed that short-ranged order causes the appearance of backfolded intensity
		even up to $T_s$, but we do not find any evidence for this}\BibitemShut
	{NoStop}%
	\bibitem [{\citenamefont {Richard}\ \emph {et~al.}(2010)\citenamefont
		{Richard}, \citenamefont {Nakayama}, \citenamefont {Sato}, \citenamefont
		{Neupane}, \citenamefont {Xu}, \citenamefont {Bowen}, \citenamefont {Chen},
		\citenamefont {Luo}, \citenamefont {Wang}, \citenamefont {Dai}, \citenamefont
		{Fang}, \citenamefont {Ding},\ and\ \citenamefont {Takahashi}}]{Richard2010}%
	\BibitemOpen
	\bibfield  {author} {\bibinfo {author} {\bibfnamefont {P.}~\bibnamefont
			{Richard}}, \bibinfo {author} {\bibfnamefont {K.}~\bibnamefont {Nakayama}},
		\bibinfo {author} {\bibfnamefont {T.}~\bibnamefont {Sato}}, \bibinfo {author}
		{\bibfnamefont {M.}~\bibnamefont {Neupane}}, \bibinfo {author} {\bibfnamefont
			{Y.-M.}\ \bibnamefont {Xu}}, \bibinfo {author} {\bibfnamefont {J.~H.}\
			\bibnamefont {Bowen}}, \bibinfo {author} {\bibfnamefont {G.~F.}\ \bibnamefont
			{Chen}}, \bibinfo {author} {\bibfnamefont {J.~L.}\ \bibnamefont {Luo}},
		\bibinfo {author} {\bibfnamefont {N.~L.}\ \bibnamefont {Wang}}, \bibinfo
		{author} {\bibfnamefont {X.}~\bibnamefont {Dai}}, \bibinfo {author}
		{\bibfnamefont {Z.}~\bibnamefont {Fang}}, \bibinfo {author} {\bibfnamefont
			{H.}~\bibnamefont {Ding}}, \ and\ \bibinfo {author} {\bibfnamefont
			{T.}~\bibnamefont {Takahashi}},\ }\bibfield  {title} {\enquote {\bibinfo
			{title} {{Observation of Dirac Cone Electronic Dispersion in
					${\mathrm{BaFe}}_{2}{\mathrm{As}}_{2}$}},}\ }\href {\doibase
		10.1103/PhysRevLett.104.137001} {\bibfield  {journal} {\bibinfo  {journal}
			{Phys. Rev. Lett.}\ }\textbf {\bibinfo {volume} {104}},\ \bibinfo {pages}
		{137001} (\bibinfo {year} {2010})}\BibitemShut {NoStop}%
	\bibitem [{\citenamefont {Ran}\ \emph {et~al.}(2009)\citenamefont {Ran},
		\citenamefont {Wang}, \citenamefont {Zhai}, \citenamefont {Vishwanath},\ and\
		\citenamefont {Lee}}]{Ran2009}%
	\BibitemOpen
	\bibfield  {author} {\bibinfo {author} {\bibfnamefont {Ying}\ \bibnamefont
			{Ran}}, \bibinfo {author} {\bibfnamefont {Fa}~\bibnamefont {Wang}}, \bibinfo
		{author} {\bibfnamefont {Hui}\ \bibnamefont {Zhai}}, \bibinfo {author}
		{\bibfnamefont {Ashvin}\ \bibnamefont {Vishwanath}}, \ and\ \bibinfo {author}
		{\bibfnamefont {Dung-Hai}\ \bibnamefont {Lee}},\ }\bibfield  {title}
	{\enquote {\bibinfo {title} {{Nodal spin density wave and band topology of
					the FeAs-based materials}},}\ }\href {\doibase 10.1103/PhysRevB.79.014505}
	{\bibfield  {journal} {\bibinfo  {journal} {Phys. Rev. B}\ }\textbf {\bibinfo
			{volume} {79}},\ \bibinfo {pages} {014505} (\bibinfo {year}
		{2009})}\BibitemShut {NoStop}%
	\bibitem [{\citenamefont {Morinari}\ \emph {et~al.}(2010)\citenamefont
		{Morinari}, \citenamefont {Kaneshita},\ and\ \citenamefont
		{Tohyama}}]{Morinari2010}%
	\BibitemOpen
	\bibfield  {author} {\bibinfo {author} {\bibfnamefont {Takao}\ \bibnamefont
			{Morinari}}, \bibinfo {author} {\bibfnamefont {Eiji}\ \bibnamefont
			{Kaneshita}}, \ and\ \bibinfo {author} {\bibfnamefont {Takami}\ \bibnamefont
			{Tohyama}},\ }\bibfield  {title} {\enquote {\bibinfo {title} {{Topological
					and Transport Properties of Dirac Fermions in an Antiferromagnetic Metallic
					Phase of Iron-Based Superconductors}},}\ }\href {\doibase
		10.1103/PhysRevLett.105.037203} {\bibfield  {journal} {\bibinfo  {journal}
			{Phys. Rev. Lett.}\ }\textbf {\bibinfo {volume} {105}},\ \bibinfo {pages}
		{037203} (\bibinfo {year} {2010})}\BibitemShut {NoStop}%
	\bibitem [{\citenamefont {Steckel}\ \emph {et~al.}(2016)\citenamefont
		{Steckel}, \citenamefont {Caglieris}, \citenamefont {Beck}, \citenamefont
		{Roslova}, \citenamefont {Bombor}, \citenamefont {Morozov}, \citenamefont
		{Wurmehl}, \citenamefont {B\"uchner},\ and\ \citenamefont
		{Hess}}]{Steckel2016}%
	\BibitemOpen
	\bibfield  {author} {\bibinfo {author} {\bibfnamefont {Frank}\ \bibnamefont
			{Steckel}}, \bibinfo {author} {\bibfnamefont {Federico}\ \bibnamefont
			{Caglieris}}, \bibinfo {author} {\bibfnamefont {Robert}\ \bibnamefont
			{Beck}}, \bibinfo {author} {\bibfnamefont {Maria}\ \bibnamefont {Roslova}},
		\bibinfo {author} {\bibfnamefont {Dirk}\ \bibnamefont {Bombor}}, \bibinfo
		{author} {\bibfnamefont {Igor}\ \bibnamefont {Morozov}}, \bibinfo {author}
		{\bibfnamefont {Sabine}\ \bibnamefont {Wurmehl}}, \bibinfo {author}
		{\bibfnamefont {Bernd}\ \bibnamefont {B\"uchner}}, \ and\ \bibinfo {author}
		{\bibfnamefont {Christian}\ \bibnamefont {Hess}},\ }\bibfield  {title}
	{\enquote {\bibinfo {title} {{Combined resistivity and Hall effect study on
					${\mathrm{NaFe}}_{1\ensuremath{-}x}{\mathrm{Rh}}_{x}\mathrm{As}$ single
					crystals}},}\ }\href {\doibase 10.1103/PhysRevB.94.184514} {\bibfield
		{journal} {\bibinfo  {journal} {Phys. Rev. B}\ }\textbf {\bibinfo {volume}
			{94}},\ \bibinfo {pages} {184514} (\bibinfo {year} {2016})}\BibitemShut
	{NoStop}%
	\bibitem [{\citenamefont {Shimojima}\ \emph {et~al.}(2014)\citenamefont
		{Shimojima}, \citenamefont {Suzuki}, \citenamefont {Sonobe}, \citenamefont
		{Nakamura}, \citenamefont {Sakano}, \citenamefont {Omachi}, \citenamefont
		{Yoshioka}, \citenamefont {Kuwata-Gonokami}, \citenamefont {Ono},
		\citenamefont {Kumigashira}, \citenamefont {B\"ohmer}, \citenamefont {Hardy},
		\citenamefont {Wolf}, \citenamefont {Meingast}, \citenamefont {L\"ohneysen},
		\citenamefont {Ikeda},\ and\ \citenamefont {Ishizaka}}]{Shimojima2014}%
	\BibitemOpen
	\bibfield  {author} {\bibinfo {author} {\bibfnamefont {T.}~\bibnamefont
			{Shimojima}}, \bibinfo {author} {\bibfnamefont {Y.}~\bibnamefont {Suzuki}},
		\bibinfo {author} {\bibfnamefont {T.}~\bibnamefont {Sonobe}}, \bibinfo
		{author} {\bibfnamefont {A.}~\bibnamefont {Nakamura}}, \bibinfo {author}
		{\bibfnamefont {M.}~\bibnamefont {Sakano}}, \bibinfo {author} {\bibfnamefont
			{J.}~\bibnamefont {Omachi}}, \bibinfo {author} {\bibfnamefont
			{K.}~\bibnamefont {Yoshioka}}, \bibinfo {author} {\bibfnamefont
			{M.}~\bibnamefont {Kuwata-Gonokami}}, \bibinfo {author} {\bibfnamefont
			{K.}~\bibnamefont {Ono}}, \bibinfo {author} {\bibfnamefont {H.}~\bibnamefont
			{Kumigashira}}, \bibinfo {author} {\bibfnamefont {A.~E.}\ \bibnamefont
			{B\"ohmer}}, \bibinfo {author} {\bibfnamefont {F.}~\bibnamefont {Hardy}},
		\bibinfo {author} {\bibfnamefont {T.}~\bibnamefont {Wolf}}, \bibinfo {author}
		{\bibfnamefont {C.}~\bibnamefont {Meingast}}, \bibinfo {author}
		{\bibfnamefont {H.~v.}\ \bibnamefont {L\"ohneysen}}, \bibinfo {author}
		{\bibfnamefont {H.}~\bibnamefont {Ikeda}}, \ and\ \bibinfo {author}
		{\bibfnamefont {K.}~\bibnamefont {Ishizaka}},\ }\bibfield  {title} {\enquote
		{\bibinfo {title} {{Lifting of xz/yz orbital degeneracy at the structural
					transition in detwinned FeSe}},}\ }\href {\doibase
		10.1103/PhysRevB.90.121111} {\bibfield  {journal} {\bibinfo  {journal} {Phys.
				Rev. B}\ }\textbf {\bibinfo {volume} {90}},\ \bibinfo {pages} {121111}
		(\bibinfo {year} {2014})}\BibitemShut {NoStop}%
	\bibitem [{\citenamefont {Fedorov}\ \emph {et~al.}(2016)\citenamefont
		{Fedorov}, \citenamefont {Yaresko}, \citenamefont {Kim}, \citenamefont
		{Kushnirenko}, \citenamefont {Haubold}, \citenamefont {Wolf}, \citenamefont
		{Hoesch}, \citenamefont {Gr{\"{u}}neis}, \citenamefont {B{\"{u}}chner},\ and\
		\citenamefont {Borisenko}}]{Fedorov2016}%
	\BibitemOpen
	\bibfield  {author} {\bibinfo {author} {\bibfnamefont {A}~\bibnamefont
			{Fedorov}}, \bibinfo {author} {\bibfnamefont {A}~\bibnamefont {Yaresko}},
		\bibinfo {author} {\bibfnamefont {T~K}\ \bibnamefont {Kim}}, \bibinfo
		{author} {\bibfnamefont {Y}~\bibnamefont {Kushnirenko}}, \bibinfo {author}
		{\bibfnamefont {E}~\bibnamefont {Haubold}}, \bibinfo {author} {\bibfnamefont
			{T}~\bibnamefont {Wolf}}, \bibinfo {author} {\bibfnamefont {M}~\bibnamefont
			{Hoesch}}, \bibinfo {author} {\bibfnamefont {A}~\bibnamefont
			{Gr{\"{u}}neis}}, \bibinfo {author} {\bibfnamefont {B}~\bibnamefont
			{B{\"{u}}chner}}, \ and\ \bibinfo {author} {\bibfnamefont {S~V}\ \bibnamefont
			{Borisenko}},\ }\bibfield  {title} {\enquote {\bibinfo {title} {{Effect of
					nematic ordering on electronic structure of FeSe}},}\ }\href
	{http://dx.doi.org/10.1038/srep36834 http://10.0.4.14/srep36834} {\bibfield
		{journal} {\bibinfo  {journal} {Scientific Reports}\ }\textbf {\bibinfo
			{volume} {6}},\ \bibinfo {pages} {36834} (\bibinfo {year}
		{2016})}\BibitemShut {NoStop}%
	\bibitem [{\citenamefont {He}\ \emph {et~al.}(2011)\citenamefont {He},
		\citenamefont {Zhang}, \citenamefont {Wang}, \citenamefont {Jiang},
		\citenamefont {Chen}, \citenamefont {Yang}, \citenamefont {Ye}, \citenamefont
		{Wu}, \citenamefont {Arita}, \citenamefont {Shimada}, \citenamefont
		{Namatame}, \citenamefont {Taniguchi}, \citenamefont {Chen}, \citenamefont
		{Xie},\ and\ \citenamefont {Feng}}]{He2011}%
	\BibitemOpen
	\bibfield  {author} {\bibinfo {author} {\bibfnamefont {C.}~\bibnamefont
			{He}}, \bibinfo {author} {\bibfnamefont {Y.}~\bibnamefont {Zhang}}, \bibinfo
		{author} {\bibfnamefont {X.F.}\ \bibnamefont {Wang}}, \bibinfo {author}
		{\bibfnamefont {J.}~\bibnamefont {Jiang}}, \bibinfo {author} {\bibfnamefont
			{F.}~\bibnamefont {Chen}}, \bibinfo {author} {\bibfnamefont {L.X.}\
			\bibnamefont {Yang}}, \bibinfo {author} {\bibfnamefont {Z.R.}\ \bibnamefont
			{Ye}}, \bibinfo {author} {\bibfnamefont {Fan}\ \bibnamefont {Wu}}, \bibinfo
		{author} {\bibfnamefont {M.}~\bibnamefont {Arita}}, \bibinfo {author}
		{\bibfnamefont {K.}~\bibnamefont {Shimada}}, \bibinfo {author} {\bibfnamefont
			{H.}~\bibnamefont {Namatame}}, \bibinfo {author} {\bibfnamefont
			{M.}~\bibnamefont {Taniguchi}}, \bibinfo {author} {\bibfnamefont {X.H.}\
			\bibnamefont {Chen}}, \bibinfo {author} {\bibfnamefont {B.P.}\ \bibnamefont
			{Xie}}, \ and\ \bibinfo {author} {\bibfnamefont {D.L.}\ \bibnamefont
			{Feng}},\ }\bibfield  {title} {\enquote {\bibinfo {title} {The orbital
				characters and $k_z$ dispersions of bands in iron-pnictide nafeas},}\ }\href
	{\doibase https://doi.org/10.1016/j.jpcs.2010.10.078} {\bibfield  {journal}
		{\bibinfo  {journal} {Journal of Physics and Chemistry of Solids}\ }\textbf
		{\bibinfo {volume} {72}},\ \bibinfo {pages} {479 -- 482} (\bibinfo {year}
		{2011})},\ \bibinfo {note} {spectroscopies in Novel Superconductors
		2010}\BibitemShut {NoStop}%
	\bibitem [{\citenamefont {Thirupathaiah}\ \emph {et~al.}(2012)\citenamefont
		{Thirupathaiah}, \citenamefont {Evtushinsky}, \citenamefont {Maletz},
		\citenamefont {Zabolotnyy}, \citenamefont {Kordyuk}, \citenamefont {Kim},
		\citenamefont {Wurmehl}, \citenamefont {Roslova}, \citenamefont {Morozov},
		\citenamefont {B\"uchner},\ and\ \citenamefont
		{Borisenko}}]{Thirupathaiah2012}%
	\BibitemOpen
	\bibfield  {author} {\bibinfo {author} {\bibfnamefont {S.}~\bibnamefont
			{Thirupathaiah}}, \bibinfo {author} {\bibfnamefont {D.~V.}\ \bibnamefont
			{Evtushinsky}}, \bibinfo {author} {\bibfnamefont {J.}~\bibnamefont {Maletz}},
		\bibinfo {author} {\bibfnamefont {V.~B.}\ \bibnamefont {Zabolotnyy}},
		\bibinfo {author} {\bibfnamefont {A.~A.}\ \bibnamefont {Kordyuk}}, \bibinfo
		{author} {\bibfnamefont {T.~K.}\ \bibnamefont {Kim}}, \bibinfo {author}
		{\bibfnamefont {S.}~\bibnamefont {Wurmehl}}, \bibinfo {author} {\bibfnamefont
			{M.}~\bibnamefont {Roslova}}, \bibinfo {author} {\bibfnamefont
			{I.}~\bibnamefont {Morozov}}, \bibinfo {author} {\bibfnamefont
			{B.}~\bibnamefont {B\"uchner}}, \ and\ \bibinfo {author} {\bibfnamefont
			{S.~V.}\ \bibnamefont {Borisenko}},\ }\bibfield  {title} {\enquote {\bibinfo
			{title} {{Weak-coupling superconductivity in electron-doped
					NaFe${}_{0.95}$Co${}_{0.05}$As revealed by ARPES}},}\ }\href {\doibase
		10.1103/PhysRevB.86.214508} {\bibfield  {journal} {\bibinfo  {journal} {Phys.
				Rev. B}\ }\textbf {\bibinfo {volume} {86}},\ \bibinfo {pages} {214508}
		(\bibinfo {year} {2012})}\BibitemShut {NoStop}%
	\bibitem [{\citenamefont {Liu}\ \emph {et~al.}(2012)\citenamefont {Liu},
		\citenamefont {Richard}, \citenamefont {Li}, \citenamefont {Jia},
		\citenamefont {Chen}, \citenamefont {Xia}, \citenamefont {Wang},
		\citenamefont {He}, \citenamefont {Yang}, \citenamefont {Pan}, \citenamefont
		{Valla}, \citenamefont {Johnson}, \citenamefont {Xu}, \citenamefont {Ding},\
		and\ \citenamefont {Wang}}]{Liu2012}%
	\BibitemOpen
	\bibfield  {author} {\bibinfo {author} {\bibfnamefont {Z.-H.}\ \bibnamefont
			{Liu}}, \bibinfo {author} {\bibfnamefont {P.}~\bibnamefont {Richard}},
		\bibinfo {author} {\bibfnamefont {Y.}~\bibnamefont {Li}}, \bibinfo {author}
		{\bibfnamefont {L.-L.}\ \bibnamefont {Jia}}, \bibinfo {author} {\bibfnamefont
			{G.-F.}\ \bibnamefont {Chen}}, \bibinfo {author} {\bibfnamefont {T.-L.}\
			\bibnamefont {Xia}}, \bibinfo {author} {\bibfnamefont {D.-M.}\ \bibnamefont
			{Wang}}, \bibinfo {author} {\bibfnamefont {J.-B.}\ \bibnamefont {He}},
		\bibinfo {author} {\bibfnamefont {H.-B.}\ \bibnamefont {Yang}}, \bibinfo
		{author} {\bibfnamefont {Z.-H.}\ \bibnamefont {Pan}}, \bibinfo {author}
		{\bibfnamefont {T.}~\bibnamefont {Valla}}, \bibinfo {author} {\bibfnamefont
			{P.~D.}\ \bibnamefont {Johnson}}, \bibinfo {author} {\bibfnamefont
			{N.}~\bibnamefont {Xu}}, \bibinfo {author} {\bibfnamefont {H.}~\bibnamefont
			{Ding}}, \ and\ \bibinfo {author} {\bibfnamefont {S.-C.}\ \bibnamefont
			{Wang}},\ }\bibfield  {title} {\enquote {\bibinfo {title} {{Orbital
					characters and near two-dimensionality of Fermi surfaces in
					NaFe1−xCoxAs}},}\ }\href {\doibase 10.1063/1.4767374} {\bibfield  {journal}
		{\bibinfo  {journal} {Applied Physics Letters}\ }\textbf {\bibinfo {volume}
			{101}},\ \bibinfo {pages} {202601} (\bibinfo {year} {2012})}\BibitemShut
	{NoStop}%
	\bibitem [{\citenamefont {Ge}\ \emph {et~al.}(2013)\citenamefont {Ge},
		\citenamefont {Ye}, \citenamefont {Xu}, \citenamefont {Zhang}, \citenamefont
		{Jiang}, \citenamefont {Xie}, \citenamefont {Song}, \citenamefont {Zhang},
		\citenamefont {Dai},\ and\ \citenamefont {Feng}}]{Ge2013}%
	\BibitemOpen
	\bibfield  {author} {\bibinfo {author} {\bibfnamefont {Q.~Q.}\ \bibnamefont
			{Ge}}, \bibinfo {author} {\bibfnamefont {Z.~R.}\ \bibnamefont {Ye}}, \bibinfo
		{author} {\bibfnamefont {M.}~\bibnamefont {Xu}}, \bibinfo {author}
		{\bibfnamefont {Y.}~\bibnamefont {Zhang}}, \bibinfo {author} {\bibfnamefont
			{J.}~\bibnamefont {Jiang}}, \bibinfo {author} {\bibfnamefont {B.~P.}\
			\bibnamefont {Xie}}, \bibinfo {author} {\bibfnamefont {Y.}~\bibnamefont
			{Song}}, \bibinfo {author} {\bibfnamefont {C.~L.}\ \bibnamefont {Zhang}},
		\bibinfo {author} {\bibfnamefont {Pengcheng}\ \bibnamefont {Dai}}, \ and\
		\bibinfo {author} {\bibfnamefont {D.~L.}\ \bibnamefont {Feng}},\ }\bibfield
	{title} {\enquote {\bibinfo {title} {{Anisotropic but Nodeless
					Superconducting Gap in the Presence of Spin-Density Wave in Iron-Pnictide
					Superconductor
					${\mathrm{NaFe}}_{1\mathbf{\ensuremath{-}}x}{\mathrm{Co}}_{x}\mathrm{As}$}},}\
	}\href {\doibase 10.1103/PhysRevX.3.011020} {\bibfield  {journal} {\bibinfo
			{journal} {Phys. Rev. X}\ }\textbf {\bibinfo {volume} {3}},\ \bibinfo {pages}
		{011020} (\bibinfo {year} {2013})}\BibitemShut {NoStop}%
	\bibitem [{\citenamefont {Watson}\ \emph {et~al.}(2015)\citenamefont {Watson},
		\citenamefont {Kim}, \citenamefont {Haghighirad}, \citenamefont {Davies},
		\citenamefont {McCollam}, \citenamefont {Narayanan}, \citenamefont {Blake},
		\citenamefont {Chen}, \citenamefont {Ghannadzadeh}, \citenamefont
		{Schofield}, \citenamefont {Hoesch}, \citenamefont {Meingast}, \citenamefont
		{Wolf},\ and\ \citenamefont {Coldea}}]{Watson2015}%
	\BibitemOpen
	\bibfield  {author} {\bibinfo {author} {\bibfnamefont {M.~D.}\ \bibnamefont
			{Watson}}, \bibinfo {author} {\bibfnamefont {T.~K.}\ \bibnamefont {Kim}},
		\bibinfo {author} {\bibfnamefont {A.~A.}\ \bibnamefont {Haghighirad}},
		\bibinfo {author} {\bibfnamefont {N.~R.}\ \bibnamefont {Davies}}, \bibinfo
		{author} {\bibfnamefont {A.}~\bibnamefont {McCollam}}, \bibinfo {author}
		{\bibfnamefont {A.}~\bibnamefont {Narayanan}}, \bibinfo {author}
		{\bibfnamefont {S.~F.}\ \bibnamefont {Blake}}, \bibinfo {author}
		{\bibfnamefont {Y.~L.}\ \bibnamefont {Chen}}, \bibinfo {author}
		{\bibfnamefont {S.}~\bibnamefont {Ghannadzadeh}}, \bibinfo {author}
		{\bibfnamefont {A.~J.}\ \bibnamefont {Schofield}}, \bibinfo {author}
		{\bibfnamefont {M.}~\bibnamefont {Hoesch}}, \bibinfo {author} {\bibfnamefont
			{C.}~\bibnamefont {Meingast}}, \bibinfo {author} {\bibfnamefont
			{T.}~\bibnamefont {Wolf}}, \ and\ \bibinfo {author} {\bibfnamefont {A.~I.}\
			\bibnamefont {Coldea}},\ }\bibfield  {title} {\enquote {\bibinfo {title}
			{{Emergence of the nematic electronic state in FeSe}},}\ }\href {\doibase
		10.1103/PhysRevB.91.155106} {\bibfield  {journal} {\bibinfo  {journal} {Phys.
				Rev. B}\ }\textbf {\bibinfo {volume} {91}},\ \bibinfo {pages} {155106}
		(\bibinfo {year} {2015})}\BibitemShut {NoStop}%
	\bibitem [{\citenamefont {Zhang}\ \emph {et~al.}(2017)\citenamefont {Zhang},
		\citenamefont {Yaji}, \citenamefont {Hashimoto}, \citenamefont {Ota},
		\citenamefont {Kondo}, \citenamefont {Okazaki}, \citenamefont {Wang},
		\citenamefont {Wen}, \citenamefont {Gu}, \citenamefont {Ding},\ and\
		\citenamefont {Shin}}]{Zhang2017_arxiv}%
	\BibitemOpen
	\bibfield  {author} {\bibinfo {author} {\bibfnamefont {P}~\bibnamefont
			{Zhang}}, \bibinfo {author} {\bibfnamefont {K}~\bibnamefont {Yaji}}, \bibinfo
		{author} {\bibfnamefont {T}~\bibnamefont {Hashimoto}}, \bibinfo {author}
		{\bibfnamefont {Y}~\bibnamefont {Ota}}, \bibinfo {author} {\bibfnamefont
			{T}~\bibnamefont {Kondo}}, \bibinfo {author} {\bibfnamefont {K}~\bibnamefont
			{Okazaki}}, \bibinfo {author} {\bibfnamefont {Z}~\bibnamefont {Wang}},
		\bibinfo {author} {\bibfnamefont {J}~\bibnamefont {Wen}}, \bibinfo {author}
		{\bibfnamefont {G.~D.}\ \bibnamefont {Gu}}, \bibinfo {author} {\bibfnamefont
			{H}~\bibnamefont {Ding}}, \ and\ \bibinfo {author} {\bibfnamefont
			{S}~\bibnamefont {Shin}},\ }\bibfield  {title} {\enquote {\bibinfo {title}
			{{Observation of topological superconductivity on the surface of iron-based
					superconductor}},}\ }\href {https://arxiv.org/abs/1706.05163} {\bibfield
		{journal} {\bibinfo  {journal} {arXiv:1706.05163}\ } (\bibinfo {year}
		{2017})}\BibitemShut {NoStop}%
	\bibitem [{\citenamefont {Wang}\ \emph {et~al.}(2015)\citenamefont {Wang},
		\citenamefont {Zhang}, \citenamefont {Xu}, \citenamefont {Zeng},
		\citenamefont {Miao}, \citenamefont {Xu}, \citenamefont {Qian}, \citenamefont
		{Weng}, \citenamefont {Richard}, \citenamefont {Fedorov}, \citenamefont
		{Ding}, \citenamefont {Dai},\ and\ \citenamefont {Fang}}]{Wang2015top}%
	\BibitemOpen
	\bibfield  {author} {\bibinfo {author} {\bibfnamefont {Zhijun}\ \bibnamefont
			{Wang}}, \bibinfo {author} {\bibfnamefont {P.}~\bibnamefont {Zhang}},
		\bibinfo {author} {\bibfnamefont {Gang}\ \bibnamefont {Xu}}, \bibinfo
		{author} {\bibfnamefont {L.~K.}\ \bibnamefont {Zeng}}, \bibinfo {author}
		{\bibfnamefont {H.}~\bibnamefont {Miao}}, \bibinfo {author} {\bibfnamefont
			{Xiaoyan}\ \bibnamefont {Xu}}, \bibinfo {author} {\bibfnamefont
			{T.}~\bibnamefont {Qian}}, \bibinfo {author} {\bibfnamefont {Hongming}\
			\bibnamefont {Weng}}, \bibinfo {author} {\bibfnamefont {P.}~\bibnamefont
			{Richard}}, \bibinfo {author} {\bibfnamefont {A.~V.}\ \bibnamefont
			{Fedorov}}, \bibinfo {author} {\bibfnamefont {H.}~\bibnamefont {Ding}},
		\bibinfo {author} {\bibfnamefont {Xi}~\bibnamefont {Dai}}, \ and\ \bibinfo
		{author} {\bibfnamefont {Zhong}\ \bibnamefont {Fang}},\ }\bibfield  {title}
	{\enquote {\bibinfo {title} {{Topological nature of the
					${\mathrm{FeSe}}_{0.5}{\mathrm{Te}}_{0.5}$ superconductor}},}\ }\href
	{\doibase 10.1103/PhysRevB.92.115119} {\bibfield  {journal} {\bibinfo
			{journal} {Phys. Rev. B}\ }\textbf {\bibinfo {volume} {92}},\ \bibinfo
		{pages} {115119} (\bibinfo {year} {2015})}\BibitemShut {NoStop}%
	\bibitem [{Note6()}]{Note6}%
	\BibitemOpen
	\bibinfo {note} {An additional rather faint outer ring of intensity is
		detected in the 27 eV data in Fig.~\ref {fig:fig-backfolding}(c). The origin
		of this feature is not clear to us but we do not believe that it is likely to
		be a bulk quasiparticle band.}\BibitemShut {Stop}%
	\bibitem [{\citenamefont {Dai}(2015)}]{Dai2015review}%
	\BibitemOpen
	\bibfield  {author} {\bibinfo {author} {\bibfnamefont {Pengcheng}\
			\bibnamefont {Dai}},\ }\bibfield  {title} {\enquote {\bibinfo {title}
			{{Antiferromagnetic order and spin dynamics in iron-based
					superconductors}},}\ }\href {\doibase 10.1103/RevModPhys.87.855} {\bibfield
		{journal} {\bibinfo  {journal} {Rev. Mod. Phys.}\ }\textbf {\bibinfo {volume}
			{87}},\ \bibinfo {pages} {855--896} (\bibinfo {year} {2015})}\BibitemShut
	{NoStop}%
	\bibitem [{\citenamefont {Ma}\ \emph {et~al.}(2011)\citenamefont {Ma},
		\citenamefont {Chen}, \citenamefont {Yao}, \citenamefont {Zhang},
		\citenamefont {Zhang}, \citenamefont {Xia},\ and\ \citenamefont
		{Yu}}]{Ma2011}%
	\BibitemOpen
	\bibfield  {author} {\bibinfo {author} {\bibfnamefont {L.}~\bibnamefont
			{Ma}}, \bibinfo {author} {\bibfnamefont {G.~F.}\ \bibnamefont {Chen}},
		\bibinfo {author} {\bibfnamefont {Dao-Xin}\ \bibnamefont {Yao}}, \bibinfo
		{author} {\bibfnamefont {J.}~\bibnamefont {Zhang}}, \bibinfo {author}
		{\bibfnamefont {S.}~\bibnamefont {Zhang}}, \bibinfo {author} {\bibfnamefont
			{T.-L.}\ \bibnamefont {Xia}}, \ and\ \bibinfo {author} {\bibfnamefont
			{Weiqiang}\ \bibnamefont {Yu}},\ }\bibfield  {title} {\enquote {\bibinfo
			{title} {{$^{23}\mathrm{Na}$ and $^{75}\mathrm{As}$ study of
					antiferromagnetism and spin fluctuations in NaFeAs single crystals}},}\
	}\href {\doibase 10.1103/PhysRevB.83.132501} {\bibfield  {journal} {\bibinfo
			{journal} {Phys. Rev. B}\ }\textbf {\bibinfo {volume} {83}},\ \bibinfo
		{pages} {132501} (\bibinfo {year} {2011})}\BibitemShut {NoStop}%
	\bibitem [{\citenamefont {Wang}\ \emph {et~al.}(2017)\citenamefont {Wang},
		\citenamefont {Park}, \citenamefont {Yu}, \citenamefont {Li}, \citenamefont
		{Song}, \citenamefont {Zhang}, \citenamefont {Ivanov}, \citenamefont
		{Kulda},\ and\ \citenamefont {Dai}}]{Wang2017}%
	\BibitemOpen
	\bibfield  {author} {\bibinfo {author} {\bibfnamefont {Weiyi}\ \bibnamefont
			{Wang}}, \bibinfo {author} {\bibfnamefont {J.~T.}\ \bibnamefont {Park}},
		\bibinfo {author} {\bibfnamefont {Rong}\ \bibnamefont {Yu}}, \bibinfo
		{author} {\bibfnamefont {Yu}~\bibnamefont {Li}}, \bibinfo {author}
		{\bibfnamefont {Yu}~\bibnamefont {Song}}, \bibinfo {author} {\bibfnamefont
			{Zongyuan}\ \bibnamefont {Zhang}}, \bibinfo {author} {\bibfnamefont
			{Alexandre}\ \bibnamefont {Ivanov}}, \bibinfo {author} {\bibfnamefont {Jiri}\
			\bibnamefont {Kulda}}, \ and\ \bibinfo {author} {\bibfnamefont {Pengcheng}\
			\bibnamefont {Dai}},\ }\bibfield  {title} {\enquote {\bibinfo {title}
			{{Orbital selective neutron spin resonance in underdoped superconducting
					${\mathrm{NaFe}}_{0.985}{\mathrm{Co}}_{0.015}\mathrm{As}$}},}\ }\href
	{\doibase 10.1103/PhysRevB.95.094519} {\bibfield  {journal} {\bibinfo
			{journal} {Phys. Rev. B}\ }\textbf {\bibinfo {volume} {95}},\ \bibinfo
		{pages} {094519} (\bibinfo {year} {2017})}\BibitemShut {NoStop}%
	\bibitem [{\citenamefont {Sprau}\ \emph {et~al.}(2017)\citenamefont {Sprau},
		\citenamefont {Kostin}, \citenamefont {Kreisel}, \citenamefont
		{B{\"{o}}hmer}, \citenamefont {Taufour}, \citenamefont {Canfield},
		\citenamefont {Mukherjee}, \citenamefont {Hirschfeld}, \citenamefont
		{Andersen},\ and\ \citenamefont {Davis}}]{Sprau2017}%
	\BibitemOpen
	\bibfield  {author} {\bibinfo {author} {\bibfnamefont {P~O}\ \bibnamefont
			{Sprau}}, \bibinfo {author} {\bibfnamefont {A}~\bibnamefont {Kostin}},
		\bibinfo {author} {\bibfnamefont {A}~\bibnamefont {Kreisel}}, \bibinfo
		{author} {\bibfnamefont {A~E}\ \bibnamefont {B{\"{o}}hmer}}, \bibinfo
		{author} {\bibfnamefont {V}~\bibnamefont {Taufour}}, \bibinfo {author}
		{\bibfnamefont {P~C}\ \bibnamefont {Canfield}}, \bibinfo {author}
		{\bibfnamefont {S}~\bibnamefont {Mukherjee}}, \bibinfo {author}
		{\bibfnamefont {P~J}\ \bibnamefont {Hirschfeld}}, \bibinfo {author}
		{\bibfnamefont {B~M}\ \bibnamefont {Andersen}}, \ and\ \bibinfo {author}
		{\bibfnamefont {J~C~S{\'{e}}amus}\ \bibnamefont {Davis}},\ }\bibfield
	{title} {\enquote {\bibinfo {title} {{Discovery of orbital-selective Cooper
					pairing in FeSe}},}\ }\href {\doibase 10.1126/science.aal1575} {\bibfield
		{journal} {\bibinfo  {journal} {Science}\ }\textbf {\bibinfo {volume}
			{357}},\ \bibinfo {pages} {75--80} (\bibinfo {year} {2017})}\BibitemShut
	{NoStop}%
	\bibitem [{\citenamefont {He}\ \emph {et~al.}(2010)\citenamefont {He},
		\citenamefont {Zhang}, \citenamefont {Xie}, \citenamefont {Wang},
		\citenamefont {Yang}, \citenamefont {Zhou}, \citenamefont {Chen},
		\citenamefont {Arita}, \citenamefont {Shimada}, \citenamefont {Namatame},
		\citenamefont {Taniguchi}, \citenamefont {Chen}, \citenamefont {Hu},\ and\
		\citenamefont {Feng}}]{He2010}%
	\BibitemOpen
	\bibfield  {author} {\bibinfo {author} {\bibfnamefont {C.}~\bibnamefont
			{He}}, \bibinfo {author} {\bibfnamefont {Y.}~\bibnamefont {Zhang}}, \bibinfo
		{author} {\bibfnamefont {B.~P.}\ \bibnamefont {Xie}}, \bibinfo {author}
		{\bibfnamefont {X.~F.}\ \bibnamefont {Wang}}, \bibinfo {author}
		{\bibfnamefont {L.~X.}\ \bibnamefont {Yang}}, \bibinfo {author}
		{\bibfnamefont {B.}~\bibnamefont {Zhou}}, \bibinfo {author} {\bibfnamefont
			{F.}~\bibnamefont {Chen}}, \bibinfo {author} {\bibfnamefont {M.}~\bibnamefont
			{Arita}}, \bibinfo {author} {\bibfnamefont {K.}~\bibnamefont {Shimada}},
		\bibinfo {author} {\bibfnamefont {H.}~\bibnamefont {Namatame}}, \bibinfo
		{author} {\bibfnamefont {M.}~\bibnamefont {Taniguchi}}, \bibinfo {author}
		{\bibfnamefont {X.~H.}\ \bibnamefont {Chen}}, \bibinfo {author}
		{\bibfnamefont {J.~P.}\ \bibnamefont {Hu}}, \ and\ \bibinfo {author}
		{\bibfnamefont {D.~L.}\ \bibnamefont {Feng}},\ }\bibfield  {title} {\enquote
		{\bibinfo {title} {{Electronic-Structure-Driven Magnetic and Structure
					Transitions in Superconducting NaFeAs Single Crystals Measured by
					Angle-Resolved Photoemission Spectroscopy}},}\ }\href {\doibase
		10.1103/PhysRevLett.105.117002} {\bibfield  {journal} {\bibinfo  {journal}
			{Phys. Rev. Lett.}\ }\textbf {\bibinfo {volume} {105}},\ \bibinfo {pages}
		{117002} (\bibinfo {year} {2010})}\BibitemShut {NoStop}%
\end{thebibliography}
% PASTED BBL FILE

\end{document}